\documentclass[11pt]{article}

\usepackage{latexsym,amsmath,amssymb,theorem,epsfig,multirow,bbm}
\usepackage[noconfig]{refstyle}
\usepackage[dvipsnames]{xcolor}
\usepackage[hyperfootnotes=false, linktocpage=true, colorlinks, citecolor=blue, linkcolor=blue, urlcolor=Maroon]{hyperref}

\usepackage{cite}

\newlength{\xtrawidth}
\newlength{\xtraheight}
\DeclareFontFamily{OT1}{rsfs10}{}
\DeclareFontShape{OT1}{rsfs10}{m}{n}{ <-> rsfs10 }{}
\DeclareMathAlphabet{\mathscript}{OT1}{rsfs10}{m}{n}

\numberwithin{equation}{section}

\newcommand{\pt}{\partial}

\def\fnote#1#2{\begingroup\def\thefootnote{#1}\footnote{#2}
     \addtocounter{footnote}{-1}\endgroup}

\def\o{\omega}

\def\cV{{\mathcal V}}

\def\be{\begin{equation}}
\def\ee{\end{equation}}
\def\bea{\begin{eqnarray}}
\def\eea{\end{eqnarray}}

\begin{document}

\begin{titlepage}

\title{{\LARGE\bf  Yukawa Unification in Heterotic String Theory}\\[1em] }
\author{
Evgeny I. Buchbinder${}^{1}$,
Andrei Constantin${}^{2}$,
James Gray${}^{3}$,
Andre Lukas${}^{4}$
}

\date{}
\maketitle
\begin{center} { 
${}^1$ {\it The University of Western Australia, \\
35 Stirling Highway, Crawley WA 6009, Australia\\[3mm]}
${}^2\,$ {\it Department of Physics and Astronomy, Uppsala University, \\ 
       SE-751 20, Uppsala, Sweden\\[3mm]}
${}^3\,$ {\it Physics Department, Robeson Hall, Virginia Tech,\\ Blacksburg, VA 24061, USA\\[3mm]}
${}^4${\it Rudolf Peierls Centre for Theoretical Physics, Oxford University,\\
1 Keble Road, Oxford, OX1 3NP, U.K.\\}}
\end{center}

\fnote{}{evgeny.buchbinder@uwa.edu.au}
\fnote{}{andrei.constantin@physics.uu.se}
\fnote{}{jamesgray@vt.edu}
\fnote{}{lukas@physics.ox.ac.uk} 

\vskip 1cm

\begin{abstract}\noindent
We analyze Yukawa unification in the the context of $E_8\times E_8$ heterotic Calabi-Yau models which rely on breaking to a GUT theory via a non-flat gauge bundle and subsequent Wilson line breaking to the standard model. Our focus is on underlying GUT theories with gauge group $SU(5)$ or $SO(10)$. We provide a detailed analysis of the fact that, in contrast to traditional field theory GUTs, the underlying GUT symmetry of these models does not enforce Yukawa unification. Using this formalism, we present various scenarios where Yukawa unification can occur as a consequence of additional symmetries. These additional symmetries arise naturally in some heterotic constructions and we present an explicit heterotic line bundle model which realizes one of these scenarios.
\end{abstract}

\thispagestyle{empty}
\end{titlepage}
\tableofcontents

\newpage
\section{Introduction}
										
One of the attractions of heterotic string phenomenology is its ability to preserve many of the successes of conventional Grand Unified Theories (GUTs) while avoiding their undesirable consequences. For example, the unification of gauge coupling constants seen in GUTs~\cite{Dimopoulos:1981yj, Dimopoulos:1981zb, Ibanez:1981yh, Einhorn:1981sx} can be reproduced in the string theory~\cite{Ginsparg:1987ee, Kaplunovsky:1987rp, Kaplunovsky:1985yy, Dixon:1990pc, Ibanez:1991zv, Ibanez:1992hc, Ibanez:1993bd, Dienes:1995bx, Dienes:1996du, Blumenhagen:2008aw}. Indeed, such unification is mandatory, barring large threshold corrections or unusual embeddings of the standard model group. This is a simple consequence of the fact that the theory only has a single gauge coupling constant in each sector at high energies~\cite{Dine:1985rz, Derendinger:1985cv}. On the other hand, it has long been known that the unification of Yukawa couplings observed in conventional Grand Unification \cite{Langacker:1980js,Ross:1985ai}, which can lead to many phenomenological issues, is not generically reproduced in heterotic models \cite{Green:1987mn}. In this paper, we will present a detailed approach to study this phenomenon in generality. This will allow us to investigate under what special circumstances (partial) Yukawa unification can, in fact, be exhibited. For some nice related work in the context of orbifold compactifications see~\cite{Kobayashi:2004ya,Kobayashi:2006wq}.

In order to address these questions it is important to be specific about the underlying class of models. For the purpose of the present paper, we will focus on the standard heterotic Calabi-Yau models with an intermediate GUT stage~\cite{Candelas:1985en,Green:1987mn}. By this we mean models which are constructed in a two-step process. In the first step, the original $E_8$ gauge group is broken to a GUT group, typically $SU(5)$ or $SO(10)$, by a gauge bundle $\hat{V}\rightarrow \hat{X}$ with a non-flat connection on a smooth Calabi-Yau manifold $\hat{X}$. In a second step, this model is divided by a discrete symmetry $\Gamma$ of $\hat{X}$ and $\hat{V}$ and the GUT group is broken to the standard model group by introducing a Wilson line on the quotient. For such constructions, a well-defined and consistent string model with GUT symmetry can be associated to the resulting standard model and we can ask if this underlying GUT model can lead to Yukawa unification. Such models can also be compared to traditional field theory GUTs with $SU(5)$ or $SO(10)$ gauge symmetry. The simplest versions of these field theory GUTs lead to the unification of d-quark and lepton Yukawa couplings for all families in the case of $SU(5)$ and to unification of all three types of Yukawa couplings in the case of $SO(10)$. Under what circumstances do heterotic Calabi-Yau models with an intermediate GUT symmetry share these properties?

At this point we may pause and ask why we insist on models with an underlying GUT symmetry (beyond the desire for unification of the gauge couplings). One may attempt instead to construct heterotic models without Wilson lines where the $E_8$ gauge group is directly broken down to the standard model group by a bundle with non-flat connection. In Refs.~\cite{Blumenhagen:2006ux,Blumenhagen:2006wj,Blumenhagen:2005ga} such models have been considered and an interesting conclusion has been obtained. First of all, it turns out it is possible to break to the standard model directly by including flux in the standard hypercharge direction within $SU(5)$ while keeping a suitable ``flipped" version of hypercharge massless and obtain standard model multiplets with the correct values of hypercharge. However, when trying to engineer a standard model spectrum within such setting a serious obstruction arises. Having broken up the spectrum into all the various standard model multiplets means that an independent index condition for each multiplet has to be imposed on the compactification. It turns out, and has been shown in Ref.~\cite{Anderson:2014hia}, that these index conditions can never be satisfied simultaneously for any Calabi-Yau. In other words, the standard model spectrum is too complicated and too fragmented to be obtained directly, at least in the heterotic context, without an underlying GUT symmetry. This problem disappears for models with an intermediate GUT symmetry and Wilson line breaking. Obtaining the correct chiral asymmetry at the GUT level requires only one index condition and the subsequent Wilson line breaking, while splitting GUT multiplets into standard model multiplets, does not change the chiral asymmetry. This strongly suggests that heterotic models with an intermediate GUT theory constitute the right approach to heterotic model building. In addition to reminding us of how non-trivial it is that the heterotic string is able to reproduce the standard model spectrum, the above comments also add emphasis to the question on what effect this intermediate GUT theory might have on the unification of Yukawa couplings. 

Returning to the main line of argument, the observation that generically Yukawa couplings do not unify is simple and relatively easy to explain by comparing the string standard model with the associated underlying GUT theory. First, consider a heterotic GUT model, with GUT group $SU(5)$ or $SO(10)$, based on a Calabi-Yau three-fold $\hat{X}$ and a vector bundle $\hat{V}\rightarrow \hat{X}$. Assume that $\hat{X}$ has a freely-acting discrete symmetry~$\Gamma$ so that $X=\hat{X}/\Gamma$ is a Calabi-Yau manifold. Further assume that the symmetry $\Gamma$ ``lifts" to the bundle $\hat{V}$ which then descends to a bundle $V\rightarrow X$ on the quotient manifold. On the quotient we add a Wilson line $W$, so that the gauge bundle becomes $V\oplus W$, in order to break the GUT symmetry to the standard model symmetry. The ``upstairs" and ``downstairs" indices are related by ${\rm ind}(V)={\rm ind}(\hat{V})/|\Gamma|$ (where $|\Gamma|$ is the order of the group $\Gamma$) while, as discussed earlier, the Wilson line does not affect the index. Hence, for a standard model with three families of quarks and leptons we require an underlying GUT model with $3|\Gamma|$ families and the associated GUT Yukawa couplings are matrices of size $(3|\Gamma|)\times (3|\Gamma|)$. It turns out, and we will show explicitly in the course of the paper, that the standard model Yukawa matrices which would be equal in the context of a field theory GUT {\it always} originate from different parts of this larger Yukawa matrices present in the GUT theory. Thus the Grand Unified symmetry itself never relates the Yukawa couplings.

For example, consider a model with $SU(5)$ GUT symmetry. For a standard field theory GUT the Yukawa couplings $Y_{ij}\,\overline{\bf 5}^H\overline{\bf 5}^i{\bf 10}^j$, where $i,j=1,2,3$, lead to $Y^{(e)}_{ij}=Y^{(d)}_{ij}$, that is, to equal lepton and d-quark Yukawa matrices. In contrast, consider a heterotic Calabi-Yau model with intermediate $SU(5)$ GUT symmetry and, say, discrete group $\Gamma=\mathbb{Z}_2$. In this case, the GUT Yukawa couplings $\hat{Y}_{IJ}\,\overline{\bf 5}^H\overline{\bf 5}^I{\bf 10}^J$, where $I,J=1,\ldots ,6$, involve six families and a $6\times 6$ Yukawa matrix $\hat{Y}_{IJ}$. One finds that the $3\times 3$ lepton and d-quark Yukawa matrices $Y^{(e)}_{ij}$ and $Y^{(d)}_{ij}$ always originate from different parts of the $6\times 6$ matrix $\hat{Y}_{IJ}$ and are, hence, unrelated by the GUT symmetry. 

The comments of the previous paragraph do not mean that Yukawa unification cannot occur in such models. For one, special choices of the upstairs Yukawa couplings $\hat{Y}_{IJ}$ can lead to Yukawa unification, although such ad-hoc choices might seem unconvincing. One might also ask whether Yukawa unification can be enforced by additional symmetries of the upstairs theory, distinct from the GUT symmetry. Having developed a concrete formalism to describe the phenomenon discussed in the previous two paragraphs, we then employ this technology to address this question. The upstairs theory is certainly invariant under the discrete symmetry $\Gamma$ and in addition, depending on the structure of the bundle $\hat{V}$, can have a number of additional $U(1)$ symmetries $\hat{J}=S(U(1)^f)$. We will show that the additional symmetries $\hat{J}$ and $\Gamma$ do not enforce Yukawa unification if they commute. On the other hand, we present scenarios with non-commuting $\hat{J}$ and $\Gamma$ which can lead to (full or partial) Yukawa unification. We also construct an explicit example, in the context of heterotic line bundle models, where such a scenario is realized. For this concrete example we compute the Yukawa couplings directly, using the formalism developed in \cite{Blesneag:2015pvz,BBL}, to demonstrate that they do not vanish and that the model does indeed exhibit unification.

In conclusion, the underlying GUT symmetry in heterotic models never enforces Yukawa unification in the same way that it does for field theory GUTs. However, in certain examples, Yukawa unification can be exhibited, being enforced by certain symmetries in the high energy theory that we characterize.

The plan of the paper is as follows. In the next section, we will review the construction of heterotic Calabi-Yau models with both underlying $SU(5)$ and $SO(10)$ GUT theories. In Section~\ref{sec:Yuk}, we analyze the relation between upstairs and downstairs Yukawa couplings and show that the GUT symmetry does not lead to unification. Scenarios where additional symmetries of the GUT theory can lead to Yukawa unification are presented in Section~\ref{sec:scen}. Section~\ref{sec:ex} provides an explicit heterotic line bundle model which realizes one of these scenarios. We conclude in Section~\ref{sec:concl}.


\section{Heterotic GUT models}\label{sec:basics}
In this section, we describe the basic model-building set-up for both $SU(5)$ and $SO(10)$ heterotic GUT models (for more details, see Refs.~\cite{Anderson:2007nc, Anderson:2008ex, Anderson:2012yf}). In either case, the ``upstairs" GUT model is based on a Calabi-Yau three-fold $\hat{X}$ with freely-acting discrete symmetry $\Gamma$ and a vector bundle $\hat{V}\rightarrow\hat{X}$ with a structure group that embeds into $E_8$ and with a $\Gamma$-equivariant structure. There is a projection $\pi:\hat{X}\rightarrow X$ to the quotient manifold $X=\hat{X}/\Gamma$ and, thanks to its equivariant structure, the bundle $\hat{V}$ descends to a bundle $V\rightarrow X$, so that $\hat{V}=\pi^* V$. The quotient manifold $X$, together with the bundle $V$ and a Wilson line $W$ on $X$ define the ``downstairs" theory. 

\subsection{Models with underlying $SU(5)$ GUT}
In this case, the structure group $\hat{H}$ of $\hat{V}$ is embedded into $E_8$ via $\hat{H}\subset SU(5)\subset E_8$, using the $SU(5)\times SU(5)$ maximal sub-group of $E_8$. The low-energy gauge group is the commutant of $\hat{H}$ with $E_8$ and we require that it is of the form $SU_{\rm GUT}(5)\times\hat{J}$, where $\hat{J}=S(U(1)^f)$ represent a certain number, $f-1$, of additional $U(1)$ symmetries. For the maximal choice of structure group, $\hat{H}=SU(5)$, there is no additional $U(1)$ symmetry and $\hat{J}$ is trivial. The other extreme is a maximally split bundle with structure group $\hat{H}=S(U(1)^5)$ for which we have four additional $U(1)$ symmetries, $\hat{J}=S(U(1)^5)$. Altogether the GUT theory has gauge symmetry $SU_{\rm GUT}(5)\times \hat{J}$ and a discrete symmetry $\Gamma$. It should be noted that the additional $U(1$) symmetries are typically Green-Schwarz anomalous and, hence, have super-heavy associated gauge bosons. 

This theory can, in principle, contain the $SU(5)$ multiplets ${\bf 10}$, $\overline{\bf 10}$, $\overline{\bf 5}$, ${\bf 5}$ and ${\bf 1}$ which are associated with the following cohomologies:
\begin{equation}
\begin{array}{rclrclrcl}
 {\bf 10}&\leftrightarrow& H^1(\hat{X},\hat{V})~,&\overline{\bf 10}&\leftrightarrow &H^1(\hat{X},\hat{V}^*)~,&
 {\bf 1}&\leftrightarrow &H^1(\hat{X},\hat{V}\otimes\hat{V}^*)~,\\[4pt]
 \overline{\bf 5}&\leftrightarrow& H^1(\hat{X},\wedge^2\hat{V})~, &
 {\bf 5}&\leftrightarrow &H^1(\hat{X},\wedge^2\hat{V}^*)~.\label{su5coh}
 \end{array}
\end{equation}

For the correct chiral asymmetry, we only have to impose the single condition ${\rm ind}(V)=-3|\Gamma|$ (since ${\rm ind}(\wedge^2V)={\rm ind}(V)$ holds in general for $SU(5)$ bundles). In order to avoid anti-families we require the absence of $\overline{\bf 10}$ multiplets, that is, $h^1(\hat{X},\hat{V}^*)=0$. Finally, we require a vector-like $\overline{\bf 5}$--${\bf 5}$ pair to account for the Higgs so we should demand that $h^1(\hat{X},\wedge^2\hat{V}^*)>0$. If these three conditions are satisfied we have obtained a GUT model with a physically promising spectrum. 

The downstair model is obtained as a quotient of the upstairs theory by $\Gamma$ and it is defined on the quotient Calabi-Yau manifold $X=\hat{X}/\Gamma$. 
Since $\hat{V}$ has a $\Gamma$-equivariant structure all cohomologies of $\hat{V}$ become $\Gamma$-representations and $\hat{V}$ descends to a bundle $V\rightarrow X$. The Wilson line $W$ on $X$ is embedded into the standard hypercharge direction within $SU_{\rm GUT}(5)$ and can be described by two characters, $\chi_2:\Gamma\rightarrow \mathbb{C}^*$ and $\chi_3:\Gamma\rightarrow\mathbb{C}^*$ satisfying $\chi_2^2\otimes\chi_3^3=1$ and $\chi_2\ncong\chi_3$. Such a Wilson line breaks $SU_{\rm GUT}(5)$ to the standard model group, $G_{\rm SM}$, so that the full downstairs gauge symmetry is $G_{\rm SM}\times J$, where $J$ is the part of $\hat{J}$ which survives the quotient. 

Since the index is unaffected by the Wilson line, the $3|\Gamma|$ upstairs families in ${\bf 10}\oplus\overline{\bf 5}$ automatically give rise to three families of quarks and leptons downstairs, with the GUT multiplets splitting in the standard way as ${\bf 10}\rightarrow (Q,u,e)$ and $\overline{\bf 5}\rightarrow (d,L)$.  The vector-like $\overline{\bf 5}$--${\bf 5}$ pairs decompose into Higgs doublets and triplets as $\overline{\bf 5}\rightarrow (T,H)$ and ${\bf 5}\rightarrow (\bar{T},\bar{H})$. For a suitable choice of equivariant structure and Wilson line it is often possible to project out the Higgs triplets and keep exactly one pair, $H,\bar{H}$, of Higgs doublets. Every downstairs multiplet acquires a Wilson line charge which is related to its hypercharge and explicitly given by
\begin{equation}
 \chi_Q=\chi_2\otimes \chi_3,\;\;\; \chi_u=\chi_3^2,\;\;\; \chi_e=\chi_2^2,\;\;\;\chi_d=\chi_3^*,\;\;\;\chi_L=\chi_2^*,\;\;\;\chi_H=\chi_2^*,\;\;\;\chi_{\bar{H}}=\chi_2\; . \label{chiSM}
\end{equation} 
Let us denote a generic downstairs multiplet by $\psi$, its associated Wilson line representation, as given above, by $\chi_\psi$ and the corresponding induced Wilson line bundle by $W_\psi$. Then, the multiplet $\psi$ is associated with the cohomologies
\begin{equation}
 \psi\;\leftrightarrow\; H^1(X,V\oplus W_\psi)\cong \left[H^1(\hat{X},\hat{V})\otimes \chi_\psi\right]_{\rm sing}\; . \label{su5sing}
\end{equation} 
The subscript ``sing" in the last expression refers to the $\Gamma$-singlets of the enclosed expression. This formula shows that the downstairs spectrum can be computed purely from representation theory of $\Gamma$ applied to the upstairs cohomology. 

\subsection{Models with underlying $Spin(10)$ GUT}
The set-up is analogous to the $SU(5)$ one. The structure group $\hat{H}$ of $\hat{V}$ is now embedded into $E_8$ via $\hat{H}\subset SU(4)\subset E_8$, using the maximal subgroup\footnote{The maximal sub-group of $E_8$ does involve $Spin(10)$, rather than $SO(10)$.} $SU(4)\times Spin(10)$ of $E_8$. Further, $\hat{H}$ should be sufficiently large such that its commutant with $E_8$ is $Spin(10)\times\hat{J}$, where $\hat{J}=S(U(1)^f)$. For the maximal choice $\hat{H}=SU(4)$ there are no additional $U(1)$ symmetries and $\hat{J}$ is trivial while the minimal choice $\hat{H}=S(U(1)^4)$ leads to three additional $U(1)$ symmetries, so $\hat{J}=S(U(1)^4)$.  Hence, the symmetry of the GUT theory includes $Spin(10)\times\hat{J}$ and the discrete symmetry $\Gamma$. As in the $SU(5)$ case, the additional $U(1)$ symmetries are typically Green-Schwarz anomalous.

The possible $Spin(10)$ multiplets in the GUT theory are ${\bf 16}$, $\overline{\bf 16}$, ${\bf 10}$ and ${\bf 1}$, with associated cohomologies
\begin{equation}
 {\bf 16}\leftrightarrow H^1(\hat{X},\hat{V}),\;\;\;\overline{\bf 16}\leftrightarrow H^1(\hat{X},\hat{V}^*),\;\;\; {\bf 10}\leftrightarrow H^1(\hat{V},\wedge^2 \hat{V}),\;\;\;{\bf 1}\leftrightarrow H^1(\hat{X},\hat{V}\otimes\hat{V}^*)\; . \label{so10coh}
\end{equation} 
As before, we need to impose one index condition, ${\rm ind}(\hat{V})=-3|\Gamma|$, for the correct chiral asymmetry. In addition, we need $h^1(\hat{X},\hat{V})=0$ for the absence of anti-families and $h^1(\hat{X},\wedge^2 \hat{V})>0$ so that at least one ${\bf 10}$ multiplet is present as a possible origin of the Higgs multiplets.

The downstairs model can be obtained by a two-step process. In the first step, we focus on the maximal sub-group $SU_{\rm GUT}(5)\times U_X(1)\subset Spin(10)$ and include a Wilson line in the $U_X(1)$ direction. Following Ref.~\cite{Braun:2004xv}, we specify the embedding of $U_X(1)=\{e^{i\theta}\;|\; \theta\in[0,2\pi]\}$ into $Spin(10)$ by using the spinor representation ${\bf 16}$. It turns out that this embedding is given by
\begin{equation}
 g_{\bf 16}(\theta)={\rm diag}(e^{-i\theta}\mathbbm{1}_{10},e^{3i\theta}\mathbbm{1}_5,e^{-5i\theta})\; , \label{g16}
\end{equation} 
which means that the branching ${\bf 16}\rightarrow {\bf 10}_{-1}\oplus \overline{\bf 5}_3\oplus{\bf 1}_{-5}$ under $SU_{\rm GUT}(5)\times U_X(1)$ provides the correctly normalized values of the $U_X(1)$ charge. We can specify the Wilson line by a character $\chi:\Gamma\rightarrow U_X(1)$ and in order for this to break $Spin(10)$ to $SU_{\rm GUT}(5)\times U_X(1)$, rather than a larger sub-group, we have to require from Eq.~\eqref{g16} that
\begin{equation}
 \chi^{-1},\; \chi^3,\; \chi^{-5}\;\mbox{are inequivalent.} \label{chiso10}
\end{equation} 
The various resulting $SU_{\rm GUT}(5)$ multiplets receive the Wilson line charges
\begin{equation}
 \chi_{\bf 10}=\chi^*,\;\;\chi_{\overline{\bf 5}}=\chi^3,\;\;\chi_{\bf 1}=\chi^{-5},\;\;\chi_{\overline{\bf 5}^H}=\chi^{-2},\;\;\chi_{{\bf 5}^{\bar{H}}}=\chi^{2}\; , \label{chiGUT}
\end{equation} 
where the last two relations follow from the branching ${\bf 10}\rightarrow {\bf 5}^{\bar{H}}_2\oplus\overline{\bf 5}^H_{-2}$ of the fundamental representation under $SU_{\rm GUT}(5)\times U_X(1)$. For the second step we can proceed as in the $SU(5)$ case and embed another Wilson line, described by two characters $\chi_2$ and $\chi_2$ with $\chi_2^2\otimes \chi_3^3=1$ and $\chi_2\ncong\chi_3$, into the standard hypercharge direction in $SU_{\rm GUT}(5)$. Under certain additional conditions on $\Gamma$, $\chi$, $\chi_2$ and $\chi_3$ which will not be relevant for our purposes the gauge group is then broken to $G_{\rm SM}\times U_{B-L}(1)\times J$, where $U_{B-L}(1)$ is obtained as a specific combination of $U_X(1)$ with hypercharge and $J$ is the part of $\hat{J}$ which survives the quotient. The standard model multiplets in this theory are characterized by their Wilson line charge $\chi_\phi$ in Eq.~\eqref{chiGUT} and the Wilson line charge $\chi_\psi$ in Eq.~\eqref{chiSM}, where $\phi={\bf 10},\overline{\bf 5},\overline{\bf 5}^H,{\bf 5}^{\bar{H}}$ and $\psi=Q,u,e,d,L,H,\bar{H}$. They are associated to the cohomologies
\begin{equation}
 (\phi,\psi)\;\leftrightarrow\; H^1(X,V\oplus W_{\phi,\psi})\cong \left[H^1(\hat{X},\hat{V})\otimes \chi_\phi\otimes \chi_\psi\right]_{\rm sing}\; .
\end{equation}


\section{Yukawa couplings upstairs and downstairs}\label{sec:Yuk}
We will now discuss Yukawa couplings in the upstairs and downstairs theories and the relation between them. Again, in order to be specific we will do this separately for $SU_{\rm GUT}(5)$ and $Spin(10)$.

\subsection{Yukawa couplings for $SU(5)$}
We begin with the Yukawa couplings in the upstairs theory. The only Yukawa couplings potentially relevant for unification are of the form $\overline{\bf 5}\,\overline{\bf 5}\,{\bf 10}$ and~\eqref{su5coh} shows that the relevant associated cohomologies are
\begin{equation}
 \cV_{\bf 10}:=H^1(\hat{X},\hat{V})\;,\quad \cV_{\overline{\bf 5}}:=H^1(\hat{X},\wedge^2\hat{V})\; . \label{V105def}
\end{equation} 
It it useful to introduce a basis
\begin{equation}
  \cV_{\bf 10}={\rm Span}\left(\nu^{({\bf 10})}_I\right)_{I=1,\ldots ,3|\Gamma|}\;,\quad \cV_{\overline{\bf 5}}={\rm Span}\left( \nu^{(\overline{\bf 5})}_I\right)_{I=1,\ldots ,3|\Gamma|+\hat{n}_H}
\end{equation}
of bundle-valued $(0,1)$-forms on these cohomologies, where we recall that we have $3|\Gamma|$ families in the upstairs theory. The number of vector-like $\overline{\bf 5}$--${\bf 5}$ pairs is denoted by $\hat{n}_H$. The four-dimensional $SU_{\rm GUT}(5)$ multiplets associated to these basis forms are denoted ${\bf 10}^I$ and $\overline{\bf 5}^I$ and the relevant Yukawa terms in the superpotential read
\begin{equation}
 \hat{W}=\hat{Y}_{IJK}\overline{\bf 5}^I\,\overline{\bf 5}^J\,{\bf 10}^K+\cdots\; ,\qquad 
 \hat{Y}_{IJK}=\int_{\hat{X}}\hat{\Omega}\wedge \nu^{(\overline{\bf 5})}_I\wedge \nu^{(\overline{\bf 5})}_J\wedge \nu^{({\bf 10})}_K\; ,
\end{equation} 
where $\hat{\Omega}$ is the holomorphic $(3,0)$-form on $\hat{X}$. This defines the holomorphic Yukawa couplings $\hat{Y}_{IJK}$ which arise in the superpotential. For the physical Yukawa couplings we also require the relevant matter field kinetic terms given by
\begin{equation}
\begin{aligned}
  \hat{K}&~=~K^{({\bf 10})}_{IJ}{\bf 10}^I{\bf 10}^{*J}~+~K^{(\overline{\bf 5})}_{IJ}\,\overline{\bf 5}^I\,\overline{\bf 5}^{*J}~+~\cdots\; ,\\[8pt]
  K^{({\bf 10})}_{IJ}&~=~\frac{1}{\hat{v}}\int_{\hat{X}}\nu^{(\bf 10)}_I\wedge *\nu^{(\bf 10)}_J\; ,\quad\quad
  K^{(\overline{\bf 5})}_{IJ}~=~\frac{1}{\hat{v}}\int_{\hat{X}}\nu^{(\overline{\bf 5})}_I\wedge *\nu^{(\overline{\bf 5})}_J\; ,
  \end{aligned}
\end{equation}  
where $\hat{v}$ is the volume of $\hat{X}$. 

The relevant multiplet types in the downstairs theory are $\psi=Q,e,d,L$, where for convenience of notation, we write the Higgs $H$ as one of the lepton doublets $L$, with associated cohomologies and basis forms
\begin{equation}
 \cV_\psi := H^1(X,V\oplus W_\psi)={\rm Span}\left(\nu^{(\psi)}_i\right)\; .
\end{equation}
The index range is $i=1,2,3$ for $\psi=Q,e,d$ and $i=1,\ldots ,3+n_H$ for $\psi=L$, where $n_H$ is the number of Higgs doublets pairs which remain from the the $\hat{n}_H$ vector-like $\overline{\bf 5}$--${\bf 5}$ pairs. The relevant superpotential and K\"ahler potential terms then read
\begin{equation}
 W=Y^{(e)}_{ijk}L^iL^je^k+Y^{(d)}_{ijk}L^id^jQ^k+\dots\; ,\qquad
 K=\sum_\psi K^{(\psi)}_{ij}\psi^i\psi^{*j}+\dots\; ,
\end{equation}
 where
 \begin{equation}
  Y^{(e)}_{ijk}=\int_X\Omega\wedge\nu^{(L)}_i\wedge \nu^{(L)}_j\wedge\nu^{(e)}_k\; ,\qquad
  Y^{(d)}_{ijk}=\int_X\Omega\wedge \nu^{(L)}_i\wedge\nu^{(d)}_j\wedge\nu^{(Q)}_k\; ,\qquad
  K^{(\psi)}_{ij}=\frac{1}{v}\int_X\nu_i^{(\psi)}\wedge *\overline{\nu}_j^{(\psi)}\; . \label{YeYd}
\end{equation}  
Here, $\Omega$ is the holomorphic $(3,0)$-form on $X$ and $v$ is the volume of $X$. 

\vspace{4pt}
We have now set up the relevant terms and couplings in both the upstairs and the downstairs theory. How are they related? The equivalence in Eq.~\eqref{su5sing} shows that the downstairs $(0,1)\text{-}\!\!$~forms~$\nu^{(\psi)}_i$, pulled back to $\hat{X}$ can be written as linear combinations of the upstairs $(0,1)\text{-}\!\!$~forms. This means
\begin{equation}
 \pi^*\nu_i^{(Q)}=c_{(Q)i}^I\nu_I^{({\bf 10})}\; ,\quad \pi^*\nu_i^{(e)}=c_{(e)i}^I\nu^{({\bf 10})}_I\;,\quad
 \pi^*\nu_i^{(d)}=c_{(d)i}^I\nu^{(\overline{\bf 5})}_I\;,\quad \pi^*\nu_i^{(L)}=c_{(L)i}^I\nu^{(\overline{\bf 5})}_I\; , \label{nurel}
\end{equation} 
where $c_{(\psi)i}^I$ are the coefficients which project onto the appropriate $\Gamma$-representations, in line with Eq.~\eqref{su5sing}. This shows that the upstairs and downstairs holomorphic Yukawa couplings are related by
\begin{equation}
 Y^{(e)}_{ijk}=c_{(L)i}^Ic_{(L)j}^Jc_{(e)k}^KY_{IJK}\; ,\qquad Y^{(d)}_{ijk}=c_{(L)i}^Ic_{(d)j}^Jc_{(Q)k}^KY_{IJK}\; , \label{Yrel'}
\end{equation} 
where $Y_{IJK}=\hat{Y}_{IJK}/|\Gamma|$.  Analogously equations
\begin{eqnarray}
K^{(Q)}_{ij}=c_{(Q)i}^I\overline{c}_{(Q)j}^JK^{(\bf 10)}_{IJ}& ,& K^{(e)}_{ij}=c_{(e)i}^I\overline{c}_{(e)j}^JK^{(\bf 10)}_{IJ}\label{KQrel}\\
K^{(d)}_{ij}=c_{(d)i}^I\overline{c}_{(d)j}^JK^{(\overline{\bf 5})}_{IJ}& ,& K^{(L)}_{ij}=c_{(L)i}^I\overline{c}_{(L)j}^JK^{(\overline{\bf 5})}_{IJ}
\label{Kdrel}
\end{eqnarray}
hold for the matter field K\"ahler metrics. Hence, the key to understanding the relation between upstairs and downstairs couplings are the coefficients $c_{(\psi)i}^I$ which, up to trivial basis transformations, are determined by $\Gamma$ representation theory. 

To make this more explicit, we recall a few simple facts from the representation theory of finite groups~\cite{Foulton:1991}. Consider a (unitary) representation $\rho:\Gamma\rightarrow {\rm Gl}({\cal V})$ of a finite group $\Gamma$ over a complex vector space ${\cal V}$. For any character $\chi:\Gamma\rightarrow \mathbb{C}^*$ we can define define the linear maps
\begin{equation}
 P_{(\chi)}=\frac{1}{|\Gamma|}\sum_{\gamma\in\Gamma}\chi(\gamma)\rho(\gamma)\; .
\end{equation} 
which project onto the sub-space of representations $\chi$ within ${\cal V}$. It is easy to verify from this definition, that the projectors for two characters $\chi$ and $\phi$ satisfy
\begin{equation}
 P_{(\chi)} P_{(\phi)} =\langle \chi,\phi\rangle P_{(\chi)}\; ,\qquad \langle \chi,\phi\rangle :=\frac{1}{|\Gamma|}\sum_{\gamma\in\Gamma}\chi^*(\gamma)\phi(\gamma)\; . \label{Porth}
\end{equation} 
In particular, this shows, setting $\phi=\chi$, that the $P_{(\chi)}$ are indeed projectors and, choosing $\chi\ncong\phi$, that $P_{(\chi)} P_{(\phi)}=0$, that is, they are orthogonal projectors provided the two characters are different. 

Returning to Yukawa couplings, we recall that, as a result of the equivariant structure on $\hat{V}$, the two relevant upstairs cohomologies $\cV_{\bf 10}$ and $\cV_{\overline{\bf 5}}$ become $\Gamma$ representations. Hence, we have two representations
\begin{equation}
 \rho_{\bf 10}:\Gamma\rightarrow {\rm Gl}(\cV_{\bf 10})\;,\quad \rho_{\overline{\bf 5}}:\Gamma\rightarrow  {\rm Gl}(\cV_{\overline{\bf 5}})\;,
\end{equation} 
which we think of as given by matrices relative to our choice of basis on each space. Given these representations we can define the projectors
\begin{equation}
\label{su5proj}
\begin{array}{lllllll}
 P_{(Q)}&=&\displaystyle\frac{1}{|\Gamma|}\sum_{\gamma\in\Gamma}\chi_Q(\gamma)\rho_{\bf 10}(\gamma)~,&\quad&
 P_{(e)}&=&\displaystyle\frac{1}{|\Gamma|}\sum_{\gamma\in\Gamma}\chi_e(\gamma)\rho_{\bf 10}(\gamma)~,\\[12pt]
 P_{(d)}&=&\displaystyle\frac{1}{|\Gamma|}\sum_{\gamma\in\Gamma}\chi_d(\gamma)\rho_{\overline{\bf 5}}(\gamma)~,&\quad&
 P_{(L)}&=&\displaystyle\frac{1}{|\Gamma|}\sum_{\gamma\in\Gamma}\chi_L(\gamma)\rho_{\overline{\bf 5}}(\gamma)~.
 \end{array}\; 
\end{equation} 
The key observation is now that the Wilson line characters $\chi_2$ and $\chi_3$ are different and, hence, from Eq.~\eqref{chiSM}, that $\chi_Q\ncong\chi_e$ and $\chi_d\ncong\chi_L$. From Eq.~\eqref{Porth} this implies orthogonality of the corresponding projectors, that is,
\begin{equation}
 P_{(Q)}P_{(e)}=0\; ,\qquad P_{(d)}P_{(L)}=0\; .
\end{equation} 
It follows that $ {\bf c}_{(Q)i}^\dagger {\bf c}_{(e)j}=(P_{(Q)}{\bf c}_{(Q)i})^\dagger (P_{(e)}{\bf c}_{(e)j})=(P_{(Q)}^2{\bf c}_{(Q)i})^\dagger (P_{(Q)}P_{(e)}{\bf c}_{(e)j})=0$ and similarly in the $d$ and $L$ sectors. 
Here ${\bf c}_{(\psi)i}$ are the vectors whose components  are $c_{(\psi)i}^I$.  As a result, we have the orthogonality relations
\begin{equation}
 {\bf c}_{(Q)i}^\dagger {\bf c}_{(e)j}=0\; ,\qquad {\bf c}_{(d)i}^\dagger {\bf c}_{(L)j}=0\;.
 \label{su5proj'}
\end{equation} 
Comparing with Eq.~\eqref{Yrel'} this means that $Y^{(e)}$ and $Y^{(d)}$ originate from different components of the upstairs Yukawa couplings $\hat{Y}$. Yukawa unification between leptons and d-quarks is still possible, provided the upstairs couplings $\hat{Y}_{IJK}$ are suitably related. However, since these couplings are independently $SU_{\rm GUT}(5)$-invariant such relations cannot be enforced but the GUT symmetry. Hence, the $SU_{\rm GUT}(5)$ symmetry does not lead to any Yukawa unification, unlike standard $SU(5)$ field theory GUTs which predict $Y^{(e)}=Y^{(d)}$.

\subsection{Yukawa couplings for $Spin(10)$}
For $Spin(10)$, the relevant superpotential term is of the form ${\bf 10}\,{\bf 16}\,{\bf 16}$ so from \eqref{so10coh} the associated cohomologies are
\begin{equation}
 \cV_{\bf 10}=H^1(\hat{X},\wedge^2\hat{V})\;,\quad \cV_{\bf 16}=H^1(\hat{X},\hat{V})\; .
\end{equation} 
As before, we introduce a basis of bundle-valued $(0,1)$-forms on these spaces 
\begin{equation}
  \cV_{\bf 10}={\rm Span}\left(\nu^{({\bf 10})}_I\right)_{A=1,\ldots ,\hat{n}_H}\;,\quad \cV_{\bf 16}={\rm Span}\left(\nu^{({\bf 16})}_I\right)_{A=1,\ldots ,3|\Gamma|}
\end{equation}
and denote the corresponding four-dimensional $Spin(10)$ multiplets by ${\bf 10}^I$ and ${\bf 16}^I$, respectively.  The relevant superpotential term is
\begin{equation}
 W=\hat{Y}_{IJK}{\bf 10}^I\,{\bf 16}^J\,{\bf 16}^K+\cdots\;,\qquad \hat{Y}_{IJK}=\int_{\hat{X}}\hat{\Omega}\wedge \nu^{({\bf 10})}_I\wedge \nu^{({\bf 16})}_J\wedge \nu^{({\bf 16})}_K\; ,
\end{equation} 
and analogous expressions for the matter field K\"ahler metrics. 

For the downstairs theory the $Spin(10)$ multiplets break up, first, into the $SU(5)$ multiplets $\phi={\bf 10},\overline{\bf 5},\overline{\bf 5}^H,{\bf 5}^{\bar{H}}$ and then into the standard model multiplets $\psi=Q,u,e,d,L,H,\bar{H}$ with associated cohomologies
\begin{equation}
 \cV_{\phi,\psi}= H^1(X,V\oplus W_{\phi,\psi})={\rm Span}\left(\nu_i^{(\psi)}\right)\; .
\end{equation}
We have the downstairs Yukawa terms
\begin{equation}
 W=Y^{(e)}_{ijk}L^iL^je^k+Y^{(d)}_{ijk}L^id^jQ^k+Y^{(u)}_{jk}\bar{H}u^jQ^k+\dots
\end{equation}
where $Y^{(e)}$ and $Y^{(d)}$ are given by Eq.~\eqref{YeYd} and 
\begin{equation}
 Y^{(u)}_{ijk}=\int_X\Omega\wedge\nu_i^{(\bar{H})}\wedge\nu_j^{(u)}\wedge\nu_k^{(Q)}\; .
\end{equation} 
With the relation of upstairs and downstairs $(0,1)$-forms as in Eq.~\eqref{nurel} the Yukawa couplings of the two theories satisfy
\begin{equation}
 Y^{(e)}_{ijk}=c_{(L)i}^Ic_{(L)j}^Jc_{(e)k}^KY_{IJK}\; ,\qquad Y^{(d)}_{ijk}=c_{(L)i}^Ic_{(d)j}^Jc_{(Q)k}^KY_{IJK}\; , \qquad
 Y^{(u)}_{ijk}=c_{(\bar{H})}^Ic_{(u)j}^Jc_{(Q)k}^KY_{IJK}\; ,
 \label{Yrel}
\end{equation} 
where $Y=\hat{Y}/|\Gamma|$. From Eqs.~\eqref{chiso10} and \eqref{chiGUT} we know that $\chi_{\bf 10}\ncong\chi_{\overline{\bf 5}}$ and a projector argument similar to the one we have used for $SU(5)$ then shows that
\begin{equation}
 {\bf c}_{(u)i}^\dagger {\bf c}_{(d)j}=0\; .
\end{equation} 
As a result, the d-quark and u-quark Yukawa matrices are not related due to the underlying GUT symmetry. Further, since $\chi_Q\ncong\chi_e$ and $\chi_d\ncong\chi_L$, the projector relations~\eqref{su5proj'} 
remain valid and imply that $Y^{(e)}$ and $Y^{(d)}$ are unrelated by the GUT symmetry. Hence our conclusion is similar to the one for $SU(5)$. In contrast to standard $SO(10)$ field theory models, 
the underlying $Spin(10)$ symmetry does not enforce any unification of the three types of Yukawa couplings. 


\section{Engineering Yukawa unification}\label{sec:scen}
In the previous section, we have seen that the underlying GUT symmetry does not lead to Yukawa unification. This happens because the downstairs Yukawa couplings which would unify in field theory GUTs originate from different parts of the underlying larger Yukawa couplings of the upstairs theory which has $3|\Gamma|$ rather than just three families. This does not mean that Yukawa unification cannot be incorporated. In particular, additional symmetry of the upstairs theory which impose relations on the upstairs Yukawa couplings might translate to unification-type relations between the downstairs Yukawa couplings. In this section, we discuss to what extent the discrete symmetry $\Gamma$ and the $U(1)$ symmetries in $\hat{J}$ may lead to such a unification. We will also present some model-building scenarios where (full or partial) unification due to these symmetries can be realised. For definiteness we will focus on models with an underlying $SU(5)$ GUT symmetry from now on, but analogous arguments can be made for $Spin(10)$. 

\subsection{A no-go statement}\label{sec:nogo}
Recall that, for models based on $SU(5)$ the two relevant cohomologies are $\cV_{\bf 10}$ and $\cV_{\overline{\bf 5}}$, as defined in Eq.~\eqref{V105def}. We have already seen that these spaces are equipped with representations of the discrete group $\Gamma$, namely
\begin{equation}
 \rho_{\bf 10}:\Gamma\rightarrow {\rm Gl}(\cV_{\bf 10})\;,\quad \rho_{\overline{\bf 5}}:\Gamma\rightarrow  {\rm Gl}(\cV_{\overline{\bf 5}})\;.
\end{equation} 
In addition, they also form representations of the $U(1)$-symmetries $\hat{J}=S(U(1)^f)$ which we denote by
\begin{equation}
R_{\bf 10}:\hat{J}\rightarrow {\rm Gl}(\cV_{\bf 10})\;,\quad R_{\overline{\bf 5}}:\hat{J}\rightarrow  {\rm Gl}(\cV_{\overline{\bf 5}})\;.
\end{equation} 
Invariance of the upstairs theory under both symmetries imposes the following conditions
\begin{equation}
 {R_{\overline{\bf 5}}(g)^L}_I{R_{\overline{\bf 5}}(g)^M}_J{R_{\bf 10}(g)^N}_KY_{LMN}=Y_{IJK}\; ,\qquad
 {\rho_{\overline{\bf 5}}(\gamma)^L}_I{\rho_{\overline{\bf 5}}(\gamma)^M}_J{\rho_{\bf 10}(\gamma)^N}_KY_{LMN}=Y_{IJK}\; . \label{inv}
\end{equation} 
on the Yukawa couplings. Our task is to translate these condition into conditions on the downstairs Yukawa couplings. We begin by writing the projectors~\eqref{su5proj} as
\begin{equation}
 P_{\psi}=\sum_i{\bf c}_{(\psi)i}^\dagger{\bf c}_{(\psi)i}\; ,
\end{equation}
where $\psi=Q,e,d,L$, assuming that the vectors ${\bf c}_{(\psi)i}$ are chosen to be orthonormal. 

Let us first discuss the implications of $\Gamma$-invariance of the upstairs Yukawa couplings. From $\rho_{\bf 10}(\gamma)P_{(\psi)}=\chi^*_\psi(\gamma)P_{(\psi)}$ (for $\psi=Q,e$) and $\rho_{\overline{\bf 5}}(\gamma)P_{(\psi)}=\chi^*_\psi(\gamma)P_{(\psi)}$ (for $\psi=d,L$) for $\gamma\in\Gamma$, it follows that
\begin{equation}
 {\rho_{\bf 10}(\gamma)^I}_Jc^J_{(\psi)j}=\chi_\psi^*(\gamma)c^I_{(\psi)j}\; ,\qquad
 {\rho_{\overline{\bf 5}}(\gamma)^I}_Jc^J_{(\psi)j}=\chi_\psi^*(\gamma)c^I_{(\psi)j}\; .
\end{equation}
Multiplying the second relation~\eqref{inv}, which expresses $\Gamma$-invariance of the upstairs Yukawa couplings, with the relevant ${\bf c}$ vectors and using the previous equations leads to 
\begin{equation}
 Y^{(e)}={\chi_L(\gamma)}^2\chi_e(\gamma)Y^{(e)}\; ,\qquad Y^{(d)}=\chi_L(\gamma)\chi_d(\gamma)\chi_Q(\gamma) Y^{(d)}\; .
\end{equation}
However, from the Wilson line relations~\eqref{chiSM} it follows immediately that 
\begin{equation}
{\chi_L(\gamma)}^2\chi_e(\gamma)=\chi_L(\gamma)\chi_d(\gamma)\chi_Q(\gamma)=1~,
\end{equation} 
so that these relations are trivially satisfied. In particular, no relations between $Y^{(e)}$ and $Y^{(d)}$ are implied.
This means, invariance of the upstairs Yukawa couplings under the discrete symmetry~$\Gamma$ does not lead to any relations between $Y^{(e)}$ and $Y^{(d)}$. 

Next, we consider the effect of the symmetry $\hat{J}$. In fact, for the purpose of our no-go statement we focus on the sub-group
\begin{equation}
 J=\{g\in\hat{J}\,|\, [R_{\bf 10}(g),\rho_{\bf 10}(\gamma)]=[R_{\overline{\bf 5}}(g),\rho_{\overline{\bf 5}}(\gamma)]=0\;\;\forall \gamma\in\Gamma\}\; , \label{Jdef}
\end{equation}
of $\hat{J}$ which commutes with $\Gamma$. For a $g\in J$ it follows immediately from the definition of the projectors~\eqref{su5proj} that $[R_{\bf 10}(g),P_{(\psi)}]=0$ for $\psi=Q,e$ and $[R_{\overline{\bf 5}}(g),P_{(\psi)}]=0$ for $\psi=d,L$. By direct calculation, this leads to
\begin{equation}
 R_{{\bf 10},K}^Ic_{(\psi),j}^K={R_{(\psi)}(g)^i}_jc_{(\psi),i}^I\; ,\qquad {R_{(\psi)}(g)^i}_j:={c^*_{(\psi),K}}^i{R_{\bf 10}(g)^K}_Jc^J_{(\psi)j}\; ,
\end{equation} 
for $\psi=Q,e$ and similarly for $R_{\overline{\bf 5}}$ and $\psi=d,L$. Then, contracting the first relation~\eqref{inv}, which reflects the $\hat{J}$-invariance of the upstairs Yukawa couplings, with the appropriate ${\bf c}$ vectors, using the previous identities and the definitions~\eqref{Yrel} of the downstairs Yukawa couplings we find
\begin{equation}
 {R_{(L)}(g)^l}_i{R_{(L)}(g)^m}_j{R_{(e)}(g)^n}_kY^{(e)}_{lmn}=Y^{(e)}_{ijk}\; ,\qquad
 {R_{(L)}(g)^l}_i{R_{(d)}(g)^m}_j{R_{(Q)}(g)^n}_kY^{(d)}_{lmn}=Y^{(d)}_{ijk}\; ,
\end{equation} 
These relations are valid for all $g\in J$ but not necessarily for all $g\in\hat{J}$.  As is evident, these relations simply reflect $J$-invariance of the downstairs theory and, while this may lead to constraints on the couplings in $Y^{(e)}$ and $Y^{(d)}$, it does not lead to unification-type relations between~$Y^{(e)}$ and~$Y^{(d)}$.

We conclude that the sub-group $J\subset \hat{J}$ which commutes with $\Gamma$ cannot cause Yukawa unification. In particular, if $J=\hat{J}$, that is, if $\hat{J}$ and $\Gamma$ commute, then neither of these symmetries can lead to unification. On the other hand, if $\hat{J}$ and $\Gamma$ do not entirely commute so that $J$ is a proper sub-group of $\hat{J}$ the non-commuting part $\hat{J}\backslash J$ of the symmetry may have some effect on Yukawa unification. This statement provides us with useful guidance for model building: We should aim to construct models where $\hat{J}$ and $\Gamma$ do not commute. In the remainder of this section, we will consider model-building scenarios with this feature and show that they can indeed lead to Yukawa unification.

\subsection{A unification scenario for $\Gamma=\mathbb{Z}_2$}\label{sec:Z2}
This scenario is within the context of heterotic line bundle models which are defined by a line bundle sum
\begin{equation}
 V=\bigoplus_{a=1}^5 L_a\; ,
\end{equation} 
with $c_1(V)=0$. For suitably generic line bundles $L_a$ the upstairs gauge symmetry of such models is $SU_{\rm GUT}(5)\times\hat{J}$, where $\hat{J}=S(U(1)^5)\cong U(1)^4$. Explicitly, we write $\hat{J}$ as
\begin{equation}
 \hat{J}=\left\{g({\boldsymbol\alpha})=(e^{i\alpha_1},\ldots ,e^{i\alpha_5})\,|\, \sum_{a=1}^5\alpha_a=0\right\}\; .
\end{equation}
For such models, the various $SU_{\rm GUT}(5)$ multiplets in the low-energy theory acquire a characteristic pattern of $S(U(1)^5)$ charges. Specifically, the ${\bf 10}$ multiplets carry charge $1$ under precisely one of the $U(1)$ symmetries and such a multiplets is denoted by ${\bf 10}_a$, where $a=1,\ldots ,5$, if it is charged under the $a^{\rm th}$ $U(1)$ symmetry. The $\overline{\bf 5}$ multiplets carry charge $1$ under precisely two $U(1)$ symmetries and are correspondingly denoted by $\overline{\bf 5}_{a,b}$, where $a,b=1,\ldots ,5$ and $a\neq b$. 

Our example is for the discrete group $\Gamma=\mathbb{Z}_2=\{1,-1\}$ and has the postulated GUT spectrum
\begin{equation}
{\cal V}_{\bf 10}={\rm Span}({\bf 10}_4,{\bf 10}_5)\;,\quad
{\cal V}_{\overline{\bf 5}}={\rm Span}( \overline{\bf 5}^H_{1,2},\overline{\bf 5}_{3,4},\overline{\bf 5}_{3,5})\; . \label{Z2spec}
\end{equation} 
(Here, we have identified the low-energy multiplets with the underlying bundle valued $(0,1)$-forms which are the actual elements of the above cohomologies.) This means that the $\hat{J}$ representations are given by
\begin{eqnarray}
 R_{\bf 10}({\boldsymbol\alpha})&=&{\rm diag}\left(e^{i{\bf e}_4\cdot{\boldsymbol\alpha}},e^{i{\bf e}_5\cdot{\boldsymbol\alpha}}\right)\label{R10ex1}\\
 R_{\overline{\bf 5}}({\boldsymbol\alpha})&=&{\rm diag}\left(e^{i({\bf e}_1+{\bf e}_2)\cdot{\boldsymbol\alpha}},
                                                                e^{i({\bf e}_3+{\bf e}_4)\cdot{\boldsymbol\alpha}},
                                                                 e^{i({\bf e}_3+{\bf e}_5)\cdot{\boldsymbol\alpha}}\right)\; ,
\end{eqnarray}
where ${\bf e}_a$ are the five-dimensional standard unit vector. For the $\mathbb{Z}_2$ representations we choose
\begin{equation}
 \rho_{\bf 10}(-1)=\sigma\;,\qquad \rho_{\overline{\bf 5}}(-1)={\rm diag}(-1,\sigma)\; ,\qquad\sigma= \left(\begin{array}{ll}0&1\\1&0\end{array}\right)\; ,
 \label{sigmaex1_1}
\end{equation} 
that is, multiplets charged under the $4^{\rm th}$ and $5^{\rm th}$ $U(1)$ symmetry are exchanged under $\mathbb{Z}_2$. Finally, we specify the Wilson by setting $\chi_2(-1)=-1$ and $\chi_3(-1)=1$ which, from Eq.~\eqref{chiSM}, implies
\begin{equation}
 \chi_Q(-1)=-1\; ,\qquad \chi_e(-1)=1\; ,\qquad \chi_d(-1)=1\;,\qquad \chi_L(-1)=-1\; , \label{Z2Wilson}
\end{equation} 
for the Wilson charges of the relevant standard model multiplets. At this point it is, of course, unclear if an actual heterotic line bundle model with all these properties can be engineered. We will see in the next section that this is, in fact, possible. For now we just proceed with the above scenario and discuss its implications for Yukawa unification.

The first observation is that the sub-group $J$ of $\hat{J}$ which commutes with $\Gamma$ (as defined in Eq.~\eqref{Jdef}) is
\begin{equation}
 J=\{g({\boldsymbol\alpha})\in\hat{J}\,|\, \alpha_4=\alpha_5\}
\end{equation}
and is, hence, a proper sub-group of $\hat{J}$. From our discussion in Section~\ref{sec:nogo} this means that there is at least a chance for Yukawa unification.
From Eqs.~\eqref{su5proj}, the projectors are easily computed as
\begin{equation}
 P_{(Q)}=\frac{1}{2}({\bf 1}_2-\sigma)\;,\qquad P_{(e)}=\frac{1}{2}({\bf 1}_2+\sigma)\; ,\qquad 
 P_{(d)}={\rm diag}(0,P_{(e)})\;,\qquad P_{(L)}={\rm diag}(1,P_{(Q)})\; . \label{projex1}
\end{equation} 
Note that the Higgs triplet is projected out (which is indicated by the zero entry in the upper left corner of  $P_{(d)}$)
while the doublet is kept (which is indicated by unity in the upper left corner of  $P_{(L)}$), as a result of choosing $\overline{\bf 5}^H_{1,2}$ to be $\mathbb{Z}_2$-odd. The corresponding ${\bf c}$ vectors are
\begin{equation}
 {\bf c}_{(Q)}=\frac{1}{\sqrt{2}}\left(\begin{array}{r}1\\-1\end{array}\right)\; ,\quad
 {\bf c}_{(e)}=\frac{1}{\sqrt{2}}\left(\begin{array}{r}1\\1\end{array}\right)\; ,\quad
 {\bf c}_{(d)}={\bf c}_{(e)}\;,\quad
 {\bf c}_{(L)}={\bf c}_{(Q)}\; ,
\end{equation} 
where, for simplicity of notation, we have left out the the Higgs direction in ${\cal V}_{\overline{\bf 5}}$.

The most general $\hat{J}$ invariant upstairs Yukawa couplings of type $\overline{\bf 5}\,\overline{\bf 5}\,{\bf 10}$ are
\begin{equation}
 \hat{W}=\overline{\bf 5}^H_{1,2}\left(\overline{\bf 5}_{3,4},\overline{\bf 5}_{3,5}\right)\hat{Y}\left(\begin{array}{l}{\bf 10}_4\\{\bf 10}_5\end{array}\right)\; ,\qquad
 \hat{Y}=2Y=\left(\begin{array}{ll}0&y\\y'&0\end{array}\right)\; . \label{Wex1}
\end{equation} 
In fact, $\Gamma=\mathbb{Z}_2$ invariance implies, in addition, that $y'=-y$ but we will not impose this for now. For the downstairs Yukawa couplings we find
\begin{equation}
 Y^{(e)}={\bf c}_{(L)}^TY{\bf c}_{(e)}={\bf c}_{(Q)}^TY{\bf c}_{(e)}=\frac{1}{4}(y-y')\; ,\quad
 Y^{(d)}={\bf c}_{(d)}^TY{\bf c}_{(Q)}={\bf c}_{(e)}^TY{\bf c}_{(Q)}=\frac{1}{4}(y'-y)\; ,
\end{equation} 
so, $Y^{(d)}=-Y^{(e)}$. The sign is, of course, physically irrelevant so that we have a case of Yukawa unification. The statement persists if we impose the $\mathbb{Z}_2$ constraint $y'=-y$ but it is, in fact, true irrespective of that. In essence, Yukawa unification in this case is a consequence of the $U(1)$ symmetry in $\hat{J}$ which does not commute with $\Gamma$ and is, hence, not contained in $J$. Under this $U(1)$ (generated by $\alpha_4-\alpha_5$) a family with subscript $4$ has charge $+1$ and a family with subscript $5$ has charge $-1$ while all other multiplets are invariant. This enforces the off-diagonal form of the Yukawa couplings in \eqref{Wex1} which, in turn, leads to Yukawa unification downstairs. The fact that $ {\bf c}_{(d)}={\bf c}_{(e)}$ and $
 {\bf c}_{(L)}={\bf c}_{(Q)}$ also means that the the matter field Kahler metrics for d and e as well as for L and Q are the same so that not only the holomorphic but also the physical Yukawa couplings unify.

This example can easily be generalized to multiple families. We can introduce $n$ pairs each of $({\bf 10}_4,{\bf 10}_5)$ and $(\overline{\bf 5}_{3,4},\overline{\bf 5}_{3,5})$ plus $(6-2n)$ families ${\bf 10}\oplus\overline{\bf 5}$ with other sets of charges, so that they cannot appear in the upstairs Yukawa couplings. Then, the above calculation goes through basically unchanged but with $y$ and $y'$ now $n\times n$ matrices. The result in the downstairs theory is Yukawa unification for $n$ families and $3-n$ families without (perturbative) Yukawa couplings.

\subsection{A unification scenario for $\Gamma=\mathbb{Z}_3$}
Following similar lines, we can also set up a scenario for the discrete group $\Gamma=\mathbb{Z}_3$, where three upstairs families are permuted. We postulate the upstairs spectrum
\begin{equation}
{\cal V}_{\bf 10}={\rm Span}({\bf 10}_3,{\bf 10}_4,{\bf 10}_5)\;,\quad
{\cal V}_{\overline{\bf 5}}={\rm Span}( \overline{\bf 5}^H_{1,2},\overline{\bf 5}_{4,5},\overline{\bf 5}_{3,5},\overline{\bf 5}_{3,4})\; ,
\end{equation} 
so that the relevant $\hat{J}$ representations are given by
\begin{eqnarray}
 R_{\bf 10}({\boldsymbol\alpha})&=&{\rm diag}\left(e^{i{\bf e}_3\cdot{\boldsymbol\alpha}},~e^{i{\bf e}_4\cdot{\boldsymbol\alpha}},~
                                                          e^{i{\bf e}_5\cdot{\boldsymbol\alpha}}\right)\\
 R_{\overline{\bf 5}}({\boldsymbol\alpha})&=&{\rm diag}\left(e^{i({\bf e}_1+{\bf e}_2)\cdot{\boldsymbol\alpha}},~
                                                                e^{i({\bf e}_4+{\bf e}_5)\cdot{\boldsymbol\alpha}},~
                                                                e^{i({\bf e}_3+{\bf e}_5)\cdot{\boldsymbol\alpha}},~
                                                                 e^{i({\bf e}_3+{\bf e}_4)\cdot{\boldsymbol\alpha}}\right)\; .
\end{eqnarray}
We write $\mathbb{Z}_3=\{1,\beta,\beta^2\}$, where $\beta=\exp(2\pi i/3)$ and introduce the representations
\begin{equation}
 \rho_{\bf 10}(\beta)=\sigma\;,\quad \rho_{\overline{\bf 5}}(\beta)={\rm diag}(1,\sigma)\;, \quad \sigma=\left(\begin{array}{lll}0&1&0\\0&0&1\\1&0&0\end{array}\right)\; .
\end{equation} 
The Wilson line is defined by $\chi_2(\beta)=1$ and $\chi_3(\beta)=\beta$ which, from Eq.~\eqref{chiSM}, leads to
\begin{equation}
 \chi_Q(\beta)=\beta\;,\quad \chi_e(\beta)=1\;,\quad \chi_d(\beta)=\beta^2\;,\quad \chi_L(\beta)=1\; .
\end{equation} 
The sub-group $J\subset \hat{J}$ which commutes with $\Gamma$ is then given by
\begin{equation}
 J=\{g({\boldsymbol\alpha})\in\hat{J}\,|\, \alpha_3=\alpha_4=\alpha_5\} \; ,
\end{equation} 
and is, hence, a proper sub-group of $\hat{J}$, as required in order to avoid the no-go statement from Section~\ref{sec:nogo}. Dropping the Higgs direction in ${\cal V}_{\overline{\bf 5}}$, we find the projectors
\begin{equation}
P_Q=\frac{1}{3}\left(\begin{array}{lll}1&\beta&\beta^2\\\beta^2&1&\beta\\\beta&\beta^2&1\end{array}\right)\; ,\quad
P_e=P_L=\frac{1}{3}\left(\begin{array}{lll}1&1&1\\1&1&1\\1&1&1\end{array}\right)\; ,\quad
P_d=\frac{1}{3}\left(\begin{array}{lll}1&\beta^2&\beta\\\beta&1&\beta^2\\\beta^2&\beta&1\end{array}\right)
\end{equation}
with associated ${\bf c}$ vectors
\begin{equation}
 {\bf c}_Q=\frac{1}{\sqrt{3}}\left(\begin{array}{l}1\\\beta^2\\\beta\end{array}\right)\; ,\quad
 {\bf c}_e={\bf c}_L=\frac{1}{\sqrt{3}}\left(\begin{array}{l}1\\1\\1\end{array}\right)\; ,\quad
 {\bf c}_d=\frac{1}{\sqrt{3}}\left(\begin{array}{l}1\\\beta\\\beta^2\end{array}\right)\; . \label{cZ3}
\end{equation}
The $\hat{J}$-invariant $\overline{\bf 5}\,\overline{\bf 5}\,{\bf 10}$ term in the superpotential reads
\begin{equation}
 \hat{W}=\overline{\bf 5}^H_{1,2}(\overline{\bf 5}_{4,5},\overline{\bf 5}_{3,5},\overline{\bf 5}_{3,4})\hat{Y}
               \left(\begin{array}{l}{\bf 10}_3\\{\bf 10}_4\\{\bf 10}_5\end{array}\right)\;,\qquad \hat{Y}=3Y={\rm diag}(\lambda_3,\lambda_4,\lambda_5)\; .
\end{equation}               
Invariance under $\Gamma$ leads to the additional constraints $\lambda_3=\lambda_4=\lambda_5$ but, as before, this is not relevant for Yukawa unification. 
For the downstairs theory this implies
\begin{equation}
Y^{(e)}={\bf c}_{(L)}^TY{\bf c}_{(e)}=\frac{1}{9}(y_3+y_4+y_5)\; ,\quad
Y^{(d)}={\bf c}_{(d)}^TY{\bf c}_{(Q)}=\frac{1}{9}(y_3+y_4+y_5)\; ,
\end{equation}
and, hence, unification of the holomorphic Yukawa couplings. Due to the $\mathbb{Z}_3$ symmetry, the upstairs K\"ahler metrics in the ${\bf 10}$ and $\overline{\bf 5}$ sectors are both proportional to the unit matrix. Even though the structure of ${\bf c}$ vectors in Eq.~\eqref{cZ3} is more complicated than in the $\mathbb{Z}_2$ case this means that the K\"ahler metrics for Q and e as well as the d and L are identical and, hence, that the physical Yukawa couplings unify as well. 

As for the $\mathbb{Z}_2$ case, we can generalize this set-up by introducing $n$ triplets $({\bf 10}_3,{\bf 10}_4,{\bf 10}_5)$ and $(\overline{\bf 5}_{4,5},\overline{\bf 5}_{3,5},\overline{\bf 5}_{3,4})$ each and $9-3n$ families ${\bf 10}\oplus\overline{\bf 5}$ with other $\hat{J}$ charges such that they cannot appear in the upstairs Yukawa couplings. Then we obtain a downstairs model with Yukawa unification for $n$ families and $3-n$ families without (perturbative) Yukawa couplings.


\section{An example with Yukawa unification}\label{sec:ex}
We would now like to construct an explicit line bundle model which realizes the $\mathbb{Z}_2$ scenario described in Section~\ref{sec:Z2}. The very specific pattern of multiplets required for this scenario imposes strong constraints on model building and it is not easy to find a viable model. In fact, our model building experience \cite{Anderson:2008uw, Anderson:2009mh, Anderson:2011ns, Anderson:2012yf, Anderson:2013xka, Buchbinder:2013dna, He:2013ofa, Buchbinder:2014qda, Buchbinder:2014sya, Buchbinder:2014qca, Constantin:2015bea} indicates that such models are quite rare, at least within the context of line bundle models. The model presented below is not realistic in that it leads to four families (starting from eight families upstairs) and contains various exotics. However, it does have a sub-sector which realizes the scenario of Section~\ref{sec:Z2} and, therefore, serves as a proof of existence.

In the first part of this section, we will present the model and show that it does indeed realize the scenario in Section~\ref{sec:Z2}. In the second part, we will compute the upstairs Yukawa couplings for this model explicitly and show that it is non-vanishing.

\subsection{The model}
The manifold underlying the model is a complete intersection Calabi-Yau (CICY)~\cite{Candelas:1987kf,Candelas:1987du} defined in the eight-dimensional ambient space ${\cal A}=\mathbb{P}^1\times\mathbb{P}^1\times\mathbb{P}^1\times\mathbb{P}^1\times\mathbb{P}^2\times\mathbb{P}^2$. Its configuration matrix reads
\begin{equation}
 \hat{X} \sim\left[\begin{array}{l|lllll}\mathbb{P}^1&1&0&1&0&0\\\mathbb{P}^1&1&0&1&0&0\\\mathbb{P}^1&0&1&0&0&1\\\mathbb{P}^1&0&1&0&0&1\\\mathbb{P}^2&1&0&0&1&1\\\mathbb{P}^2&0&1&1&1&0\end{array}\right]^{6,26}_{-40}\; . \label{cicy4109}
\end{equation} 
Here, the column vectors, which we also denote by ${\bf q}_r$, where $r=1,\ldots ,5$, represent the multi-degrees of five polynomials whose common zero locus in ${\cal A}$ defines the Calabi-Yau manifold 
$\hat{X}$. The superscript in Eq.~\eqref{cicy4109} gives the Hodge numbers $h^{1,1}(\hat{X}), h^{2,1}(\hat{X})$ and the subscript corresponds to the Euler 
number of $\hat{X}$. It will be useful to introduce the line bundles ${\cal N}_r={\cal O}_{\cal A}({\bf q}_r)$ whose sections are the defining polynomials, as well as their sum
\begin{equation}
 {\cal N}=\bigoplus_{r=1}^5{\cal N}_r\; 
\end{equation} 
whose restriction $N={\cal N}|_X$ is the normal bundle of $\hat{X}$ in ${\cal A}$. 
We also denote the homogeneous coordinates of the four $\mathbb{P}^1$ factors by $x_{i,\alpha}$, where $i=1,2,3,4$ and $\alpha=0,1$ and the homogenous coordinates of the two $\mathbb{P}^2$ factors by ${\bf y}=(y_0,y_1,y_2)^T$ and ${\bf z}=(z_0,z_1,z_2)^T$. 

For suitably restricted defining polynomials, this manifold has a freely-acting $\mathbb{Z}_2$ symmetry~\cite{Braun:2010vc} which acts on the homogeneous coordinates as
\begin{equation}
 x_{i,\alpha}\rightarrow (-1)^{\alpha+1}x_{i,\alpha}\;,\quad {\bf y}\leftrightarrow{\bf z}\; , \label{ctrans}
\end{equation} 
and on the defining equations or, equivalently, the line bundles ${\cal N}_r$ as
\begin{equation}
 {\cal N}_1\leftrightarrow {\cal N}_3\;,\quad  {\cal N}_2\leftrightarrow {\cal N}_5\;,\quad {\cal N}_4\rightarrow{\cal N}_4\; . \label{Ntrans}
\end{equation} 

The line bundle model is defined by a sum of five line bundles ${\cal L}_a\rightarrow {\cal A}$ and their restrictions $L_a={\cal L}_a|_{\hat{X}}$ to $\hat{X}$ which are explicitly given by
\begin{equation}
\begin{array}{lll}
 L_1={\cal O}_{\hat{X}}(-1,0,-1,1,0,0)\;,&L_2={\cal O}_{\hat{X}}(2,1,2,0,-1,-1)\;,&L_3={\cal O}_{\hat{X}}(1,1,-1,-1,0,0)\; ,\\[4pt]
 L_4={\cal O}_{\hat{X}}(-1,-1,0,0,0,1)\;,&L_5={\cal O}_{\hat{X}}(-1,-1,0,0,1,0)\; .&
\end{array} 
\end{equation} 
Using the methods developed in Refs.~\cite{Hubsch:1992nu, Anderson:2008ex, Anderson:2013qca, cicypackage} the line bundle cohomology of $L_a$ and their tensor product can be calculated as
\begin{equation}
 \begin{array}{rrrrrr}
 h^\bullet(\hat{X}, L_2)&=&(0,6,0,0)\;,&h^\bullet(\hat{X}, L_4)&=&(0,1,0,0)\; , \\
 h^\bullet(\hat{X}, L_5)&=&(0,1,0,0)\; ,& ~~~~~~~h^\bullet(\hat{X}, L_1\otimes L_2)&=&(0,2,0,0)\;,\\
 h^\bullet(\hat{X}, L_1\otimes L_3)&=&(0,2,0,0)\;,&h^\bullet(\hat{X}, L_1\otimes L_4)&=&(0,0,2,0)\;,\\
 h^\bullet(\hat{X},L_1\otimes L_5)&=&(0,0,2,0)\;,&h^\bullet(\hat{X},L_3\otimes L_4)&=&(0,1,0,0)\;,\\
 h^\bullet(\hat{X},L_3\otimes L_5)&=&(0,1,0,0)\;,&h^\bullet(\hat{X},L_4\otimes L_5)&=&(0,7,1,0)\; ,
 \end{array}
\end{equation}  
with all other cohomologies of $L_a$ and $L_a\otimes L_b$ appearing in wedge products of the sum of line bundles vanishing. These results can be translated into the GUT spectrum
\begin{equation}
\begin{array}{llllll}
 {\bf 10}_4\;,&{\bf 10}_5\;,&2\,\overline{\bf 5}_{1,2}^H\;,&\overline{\bf 5}_{3,4}\; ,&\overline{\bf 5}_{3,5}&\\
 6\,{\bf 10}_2\;,&2\,{\overline {\bf 5}}_{1,3}\;,&2\,{\bf 5}_{1,4}\;,&2\,{\bf 5}_{1,5}\;,&7\,\overline{\bf 5}_{4,5}\;,&{\bf 5}_{4,5}\; .
\end{array}\label{exspec}
\end{equation} 
Comparison with Eq.~\eqref{Z2spec} shows that, apart from the presence of two rather than one Higgs multiplet, the top line realizes the spectrum required for the $\mathbb{Z}_2$ unification scenario for one family while the remainder of the spectrum in the bottom line accounts for three more families and some exotics. Clearly, this model is not realistic but does contain a sub-sector of the required type on which we focus. Of course we still have to check that the multiplets in this sub-sector have the correct $\mathbb{Z}_2$ transformation properties. To this end, we determine the cohomologies for the multiplets in the first line of the spectrum~\eqref{exspec} more explicitly. By chasing through the relevant Koszul sequences we learn that these cohomologies can be expressed in terms of ambient space cohomologies as follows.
\begin{eqnarray}
 H^1(\hat{X},L_a)&\cong&H^2({\cal A},{\cal N}^*\otimes{\cal L}_a)=H^2({\cal A},{\cal N}_b^*\otimes {\cal L}_a)=H^2({\cal A},{\cal O}_{\cal A}(-2,-2,0,0,0,0))\nonumber\\
 &\cong&{\rm Span}\left(\frac{1}{x_{1,0}\,x_{1,1}\,x_{2,0}\,x_{2,1}}\right)\;\mbox{for}\; (a,b)=(4,3),(5,1)\label{HL4L5}\\[8pt]
  H^1(\hat{X},L_3\otimes L_a)&\cong&H^2({\cal A},{\cal N}^*{\otimes}{\cal L}_3{\otimes} {\cal L}_a)=H^2({\cal A},{\cal N}_b^*{\otimes} {\cal L}_3{\otimes}{\cal L}_a)=H^2({\cal A},{\cal O}_{\cal A}(0,0,{-}2,{-}2,0,0))\nonumber\\
 &\cong&{\rm Span}\left(\frac{1}{x_{3,0}\,x_{3,1}\,x_{4,0}\,x_{4,1}}\right)\;\mbox{for}\; (a,b)=(4,2),(5,5) \label{LL1}\\[8pt]
 H^1(\hat{X},L_1\otimes L_2)&\cong&H^4(\wedge^3{\cal N}^*\otimes {\cal L}_1\otimes{\cal L}_2)\nonumber\\
 &=&H^4({\cal A},{\cal N}_1^*\otimes{\cal N}_2^*\otimes{\cal N}_4^*\otimes {\cal L}_1\otimes{\cal L}_2)\oplus
 H^4({\cal A},{\cal N}_3^*\otimes{\cal N}_4^*\otimes{\cal N}_5^*\otimes {\cal L}_1\otimes{\cal L}_2)\nonumber\\
 &=&H^4({\cal A},{\cal O}_{\cal A}(0,0,0,0,-3,-3))^{\oplus 2}\nonumber\\
 &\cong&{\rm Span}\left(\frac{1}{y_0\,y_1\,y_2\,z_0\,z_1\,z_2}\right)^{\oplus 2} \;.
\label{LL2}
\end{eqnarray}
These results, together with the transformations~\eqref{ctrans} and \eqref{Ntrans}, can be used to determine the $\mathbb{Z}_2$ representations of the various multiplets. For example, the explicit representation for $H^1(\hat{X}, L_4)$ and $H^1(\hat{X}, L_5)$ in terms of coordinates show that these cohomologies are invariant under the coordinate part~\eqref{ctrans} of the $\mathbb{Z}_2$ action. 
However, the action~\eqref{Ntrans} on the normal bundle exchanges ${\cal N}_1$ and ${\cal N}_3$ which means, again from Eq.~\eqref{HL4L5} that the multiplets ${\bf 10}_4$ and ${\bf 10}_5$ are exchanged. The transformation of the other multiplets can be reasoned out in a similar fashion. The end result is that the cohomologies
\begin{equation}
 \cV_{\bf 10}={\rm Span}\left({\bf 10}_4,{\bf 10}_5\right)\;,\qquad \cV_{\overline{\bf 5}}={\rm Span}\left(\overline{\bf 5}_{1,2}^{H,1},\overline{\bf 5}_{1,2}^{H,2},\overline{\bf 5}_{3,4}\overline{\bf 5}_{3,5}\right)
\end{equation} 
carry the $\mathbb{Z}_2$ representations
\begin{equation}
 \rho_{\bf 10}(-1)=\sigma=\left(\begin{array}{cc}0&1\\1&0\end{array}\right)\;,\qquad \rho_{\overline{\bf 5}}(-1)={\rm diag}(\sigma,\sigma)\; .
\end{equation} 
This differs from the required transformation~\eqref{sigmaex1_1} only in that two Higgs multiplets are present. We can get to a complete match by focusing on the $\mathbb{Z}_2$ odd combination $\overline{\bf 5}_{1,2}^H=\overline{\bf 5}_{1,2}^{H,1}-\overline{\bf 5}_{1,2}^{H,2}$. Then, using the same Wilson line choice as in Eq.~\eqref{Z2Wilson} will project out the Higgs triplet from this odd combination and keep the Higgs doublet. (For the even combination it is, of course, the other way around and the Higgs triplet will be kept.) Focusing on this sub-sector we have indeed the same Yukawa couplings of type $\overline{\bf 5}\,\overline{\bf 5}\,{\bf 10}$ as in Eq.~\eqref{Wex1}, namely
\begin{equation}
 \hat{W}=\overline{\bf 5}^H_{1,2}\left(\overline{\bf 5}_{3,4},\overline{\bf 5}_{3,5}\right)\hat{Y}\left(\begin{array}{l}{\bf 10}_4\\{\bf 10}_5\end{array}\right)\; ,\qquad
 \hat{Y}=2Y=\left(\begin{array}{ll}0&y\\y'&0\end{array}\right)\; . \label{Wex}
\end{equation} 
This leads to Yukawa unification for one family of d-quarks and leptons from the arguments presented in Section~\ref{sec:Z2}.


\subsection{Explicit computation of the Yukawa coupling}
While our previous example realizes the correct multiplet structure required for the Yukawa unification scenario it is of course important for any meaningful statement about Yukawa unification that the requisite Yukawa couplings in~\eqref{Wex} are indeed non-zero. There are no obvious symmetry reasons to forbid these couplings but, as has been observed in Refs.~\cite{Anderson:2010tc,Blesneag:2015pvz, BBL}, there may be other reasons for the absence of perturbative Yukawa couplings in string theory. Following the methods developed in Refs.~\cite{Blesneag:2015pvz, BBL}, we will now explicitly calculate the Yukawa couplings in 
Eq.~\eqref{Wex} and show that they are non-zero.

The upstairs Yukawa couplings are given by the following general expression 
\be 
\hat{\lambda}_{IJK} = \int_{\hat{X}}
\Omega   \wedge  \nu^{(H^I)}  \wedge  \nu^{(\overline{\bf 5}^J)} \wedge \nu^{({\bf 10}^K)}\;. 
\label{e1}
\ee
We would like to compute the Yukawa couplings for the particles in the first line of the spectrum~\eqref{exspec}, so that $I, J,K =1, 2$.
We  denote $\overline{\bf 5}^1= \overline{\bf 5}_{3,4}$, $\overline{\bf 5}^2= \overline{\bf 5}_{3,5}$, ${\bf 10}^1= {\bf 10}_5$, ${\bf 10}^2= {\bf 10}_4$
and $H^1, H^2$ represent the two Higgs fields $\overline{\bf 5}^{H}_{1,2}$ in~\eqref{exspec}.
According to the computational procedure developed in Refs.~\cite{Blesneag:2015pvz, BBL} we can lift the integral to the ambient space as%
\be 
\hat{\lambda}_{IJK} = \frac{1}{ (2 \pi i)^5} \int_{{\cal A}}\mu \wedge \hat{\nu}^{(H^I)}  \wedge  \hat{\nu}^{(\overline{\bf 5}^J)} \wedge \hat{\nu}^{({\bf 10}^K)}
\wedge {\overline \pt} \Big(\frac{1}{p_1} \Big)\wedge \dots \wedge {\overline \pt} \Big(\frac{1}{p_5} \Big)\;. 
\label{e2}
\ee
Here $\hat{\nu}$ for each particle is the lift of the corresponding form $\nu$ from $\hat{X}$ to ${\cal A}$, that is $\nu =\hat{\nu}|_{\hat{X}}$, 
$p_1, \dots, p_5$ are the defining polynomials described in~\eqref{cicy4109} and $\mu$ is the holomorphic volume form 
on the ambient space. On a single projective space ${\mathbb P}^n$ with homogeneous coordinates $x_i$, $\mu$ is given by 
\be 
\mu= \frac{1}{n!} x_{a_0} d x_{a_1} \wedge \dots \wedge d x_{a_n} \epsilon^{a_0 a_1 \dots a_n}~,
\label{e3}
\ee
while on a product of projective spaces $\mu$ is given by the wedge product of the individual holomorphic volume forms on each projective space. Since an integral over ${\mathbb P}^n$  
can be viewed as an integral over ${\mathbb C}^n$ (provided all the forms are well-defined as forms on ${\mathbb P}^n$) we can introduce affine coordinates $w$ on ${\mathbb P}^n$
in which $\mu$ is simply 
\be 
\mu=  d^n w\;.
\label{e4}
\ee
In the present case we have 
\be 
\mu = d w_1 \wedge  d w_2 \wedge  d w_3 \wedge dw_4 \wedge du_1 \wedge du_2\wedge dv_1 \wedge dv_2\;,
\label{e5}
\ee
where $w_i$ are  affine coordinates on the four ${\mathbb P}^1$ spaces in~\eqref{cicy4109}, and $u_i, v_i$ are affine coordinates on the two ${\mathbb P}^2$ spaces in~\eqref{cicy4109}.

As explained in~\cite{Blesneag:2015pvz, BBL} the forms $\hat{\nu}$ are, in general, no longer closed. However, they are related to a collection of 
closed forms on ${\cal A}$ which can be obtained using the Koszul exact sequence and the corresponding cohomology long exact sequence. 
Let $\nu \in H^1 ({\hat X}, K)$ for some line bundle $K$ and $\hat{\nu} \in \Omega^1 ({\cal A}, {\cal K})$ where $K= {\cal K}|_{\hat{X}}$. If $\hat{X}$ is of 
co-dimension $m$ in ${\cal A}$ the Koszul sequence has the form 
\be 
0 \longrightarrow \Lambda^m {\cal N}^* \otimes {\cal K} \stackrel{q_{m-1}}{\longrightarrow} \Lambda^{m-1} {\cal N}^* \otimes {\cal K} \stackrel{q_{m-2}}{\longrightarrow}
\dots \stackrel{q_1}{\longrightarrow} {\cal N}^*\otimes {\cal K}  \stackrel{p}{\longrightarrow}  {\cal K}  \stackrel{r}{\longrightarrow} K \to 0\,. 
\label{e6}
\ee
Here ${\cal N}$ is the normal bundle, $r$ is the restriction map, $p= (p_1, \dots, p_m)$ is the row vector of $m$ defining polynomials 
and $q_a$ are the induced maps between higher exterior powers of vector bundles. The maps $q_a$ are uniquely fixed (up to a constant which can be absorbed 
in the coefficients of the polynomials $p_a$)
by the composition properties 
\be 
q_a \circ q_{a+1}\,, \quad p \circ q_1 =0
\label{e7}
\ee
and by the degrees of the vector bundles in~\eqref{e6}. It was shown in~\cite{Blesneag:2015pvz, BBL} that $\hat{\nu}$ is obtained by solving 
the following system of differential equations
\bea
&&
\nu =\hat{\nu}|_{\hat{X}} \,,     \qquad   \qquad \qquad \qquad \ \ \nu \in H^1(\hat{X}, K)\,, \nonumber \\
&&
{\overline \pt} \hat{\nu} = p  \hat{\omega}_1 \,,    \qquad   \qquad \qquad  \quad  \ \ \   \hat{\nu} \in \Omega^1 ({\cal A}, {\cal K})\,, \nonumber  \\
&&
\overline \pt \hat{\o}_1   = q_1 \hat{\o}_2\,,    \qquad   \qquad \qquad  \ \ \ \hat{\o}_1 \in \Omega^2 ({\cal A}, {\cal N}^* \otimes {\cal K})\,, \nonumber \\
&&
\dots \qquad \dots \qquad \dots \qquad \dots
\nonumber \\
&&
\overline \pt \hat{\o}_{m-1}   = q_{m-1} \hat{\o}_m\,,    \qquad   \qquad   \hat{\o}_{m-1} 
\in \Omega^m ({\cal A}, \Lambda^{m-1}{\cal N}^* \otimes {\cal K})\,, \nonumber \\
&&
\overline \pt \hat{\o}_m =0 \,,   \qquad   \qquad \qquad \quad \ \ \  \hat{\o}_m \in H^{m+1} ({\cal A}, \Lambda^m{\cal N}^* \otimes {\cal K})\,.
\label{e8}
\eea
The consistency of this system follows from~\eqref{e7}. The general solution to~\eqref{e8} is given by the general solution 
to the homogeneous equations and a partial solution to the inhomogeneous ones. The former describes closed forms, 
that is elements in $H^{k+1} ({\cal A}, \Lambda^k {\cal N}^* \otimes {\cal K}), k=0, 1, \dots, m-1$. The total number of  independent closed forms 
obtained this way is in one-to-one correspondence with the number of particles described by $\nu$.\footnote{There is a subtlety that, in general, thus obtained closed forms 
do not span the entire  space $H^{k+1} ({\cal A}, \Lambda^k {\cal N}^* \otimes {\cal K})$ but rather a subspace in it given
by the kernel or cokernel of  $q_a$. This all can be obtained from the cohomology long exact sequence corresponding to the Koszul sequence~\eqref{e6}.
This subtlety will not play any role in the present example and we will not discuss it. See Refs.~\cite{Blesneag:2015pvz, BBL} 
for more details.}
Since Yukawa couplings depend 
only on the cohomology classes we can choose these closed forms to be harmonic forms on ${\cal A}$ with respect to the Fubini-Study metric. 
Such forms were explicitly constructed in~\cite{Blesneag:2015pvz, BBL}.
Knowing the general solution to the homogeneous equation we can then solve~\eqref{e8} to find a partial solution to the inhomogeneous equations. 
Of course, in some cases it may happen that the system~\eqref{e8} is truncated at a earlier step and 
$\hat{\o}_s= \hat{\o}_{s+1}= \dots = \hat{\o}_{m}=0$ for some $s \leq m$. Then the maximal degree of  an ambient space closed form associated to $\nu$ is $s < m+1$. 

Let us now apply this procedure to the present example. All relevant non-vanishing cohomology groups with coefficients in 
$\Lambda^k {\cal N}^* \otimes {\cal K}$ for relevant ${\cal K}$ are given in Eqs.~\eqref{HL4L5}, \eqref{LL1}, \eqref{LL2}. 
Let us start with $\hat{\nu}^{(H^I)}$. We see that they are related to closed $(0,4)$-forms. 
Hence, the system of equations becomes
\begin{equation}\label{e9}
{\overline \pt} \hat{\nu}^{(H^I)} = p  \hat{\omega}_1^{(H^I)} \,, \qquad
\overline \pt \hat{\o}_1^{(H^I)}   = q_1 \hat{\o}_2^{(H^I)}\,, \qquad
\overline \pt \hat{\o}_2^{(H^I)}   = q_2 \hat{\o}_3^{(H^I)}\,, \qquad
\overline \pt \hat{\o}_3^{(H^I)} =0\,. 
\end{equation}
Since $H^{k+1} ({\cal A}, \Lambda^k {\cal N}^* \otimes {\cal L}_1 \otimes  {\cal L}_2)=0$ for all $k$ except $k=3$, it follows that the only solution to the homogeneous system 
is  $\hat{\o}_3^{(H^I)} $ and all partial solutions to the inhomogeneous equations are restored using $\hat{\o}_3^{(H^I)} $.
Since the form $\hat{\o}_3^{(H^I)} $ takes values in  $\Lambda^3 {\cal N}^* \otimes {\cal L}_1 \otimes  {\cal L}_2$ it can be viewed as a  
tensor or rank 3 whose components we will denote as $\hat{\o}_{3, abc}^{(H^I)}$, where 
\be 
\hat{\o}_{3, abc}^{(H^I)} \in H^{4} ({\cal A},  {\cal N}^*_a \otimes {\cal N}^*_b \otimes {\cal N}^*_c   \otimes {\cal L}_1 \otimes  {\cal L}_2)\;, \quad a<b<c\,. 
\label{e10.0}
\ee
From eq.~\eqref{LL2} we see that $\hat{\o}_{3, abc}^{(H^I)}$ has only following non-vanishing components
\be
\hat{\o}_{3, 3 4 5}^{(H^1)}\;, \quad  \hat{\o}_{3, 1 2 4}^{(H^2)}\;.
\label{e10}
\ee
Using the results of~\cite{Blesneag:2015pvz, BBL} we can write down harmonic representatives of these forms. Since 
\bea
H^{4} ({\cal A}, {\cal N}^*_1 \otimes {\cal N}^*_2 \otimes {\cal N}^*_4 \otimes {\cal L}_1 \otimes  {\cal L}_2) &= &H^{4} ({\cal A}, {\cal N}^*_3 \otimes {\cal N}^*_4 \otimes {\cal N}^*_5 \otimes {\cal L}_1 \otimes  {\cal L}_2)
\nonumber \\
&=&H^{4} ({\cal A}, {\cal O}_{{\cal A}} (0, 0, 0, 0, -3, -3))\;, \nonumber \\ 
h^{4} ({\cal A}, {\cal O}_{{\cal A}} (0, 0, 0, 0, -3, -3)) &= &1
\label{e11}
\eea
it follows that  $\hat{\o}_{3, 3 4 5}^{(H^1)}$ and  $\hat{\o}_{3, 1 2 4}^{(H^2)}$ are equal to each up to a coefficient. Their harmonic representatives are 
\bea
&& 
\hat{\o}_{3, 3 4 5}^{(H^1)} = \frac{a_1}{ (1+ |u_1|^2 + |u_2|^2)^3 (1+ |v_1|^2 + |v_2|^2)^3 } d{\overline u}_1 \wedge d{\overline u}_2 \wedge d{\overline v}_1 \wedge d{\overline v}_2\;, 
\nonumber \\
&& 
\hat{\o}_{3, 1 2 4}^{(H^2)} = \frac{a_2}{ (1+ |u_1|^2 + |u_2|^2)^3 (1+ |v_1|^2 + |v_2|^2)^3 } d{\overline u}_1 \wedge d{\overline u}_2 \wedge d{\overline v}_1 \wedge d{\overline v}_2\;. 
\label{e12}
\eea
The coefficients $a_1$ and $a_2$ can be absorbed into the four-dimensional field $H^1$ and $H^2$. 
However, we will keep them for reasons that will become clear later on. 
The solution for the lower-degree forms (which are not closed) in~\eqref{e9} can be obtained using the explicit form of $\hat{\o}_{3, abc}^{(H^I)}$
in~\eqref{e12} and the maps $p, q_1, q_2$. It is a very lengthy calculation but fortunately these forms will not be needed.

Now we apply the system~\eqref{e8} to $\hat{\nu}^{({\overline {\bf 5}^I})}$. We see from Eq.~\eqref{LL1} that it is associated to a closed $(0,2)$-from and the system~\eqref{e8} becomes
\bea
&&
{\overline \pt} \hat{\nu}^{({\overline {\bf 5}^I})} = 
p  \hat{\omega}_1^{({\overline {\bf 5}^I})} \,,  
\nonumber \\
&&
\overline \pt \hat{\o}_1^{({\overline {\bf 5}^I})}  = 0\,.    
\label{e13}
\eea
Since all $H^{k+1} ({\cal A}, \Lambda^k {\cal N}^* \otimes {\cal L}_3 \otimes  {\cal L}_a)=0$ except for $k=1$ it follows that the only solution to the homogeneous system 
is  $\hat{\o}_1^{({\overline {\bf 5}^I})} $ and a partial solution for $\hat{\nu}^{({\overline {\bf 5}^I})}$ is obtained using $\hat{\o}_1^{({\overline {\bf 5}^I})} $ by solving~\eqref{e13}. 
We view the forms $\hat{\o}_1^{({\overline {\bf 5}^I})}$ as (column) vectors with components $\hat{\o}_{1, a}^{({\overline {\bf 5}^I})} \in H^{2} ({\cal A}, {\cal N}_a^* \otimes {\cal L}_3 \otimes  {\cal L}_b)$, 
where the index $b$ labels  different $\overline {\bf 5}$ multiplets just like the index $I$. From Eq.~\eqref{LL1} we see that  $\hat{\o}_1^{({\overline {\bf 5}^I})}$ have the following non-vanishing components
\be
\hat{\o}_{1, 2}^{({\overline {\bf 5}^1})} \in H^2 ({\cal A}, {\cal N}_2^* \otimes {\cal L}_3 \otimes  {\cal L}_4)\;, \quad 
\hat{\o}_{1, 5}^{({\overline {\bf 5}^2})} \in H^2 ({\cal A}, {\cal N}_5^* \otimes {\cal L}_3 \otimes  {\cal L}_5)\;.
\label{e14}
\ee
Since
\bea
&&
H^{2} ({\cal A}, {\cal N}^*_5  \otimes {\cal L}_3 \otimes  {\cal L}_5) = H^{2} ({\cal A}, {\cal N}^*_2 \otimes {\cal L}_3 \otimes  {\cal L}_4)
=H^{2} ({\cal A}, {\cal O}_{{\cal A}} (0, 0, -2, -2, 0, 0))\;, \nonumber \\
&& 
h^{2} ({\cal A}, {\cal O}_{{\cal A}} (0, 0, -2, -2, 0, 0))=1
\label{e15.0}
\eea
it follows that the forms in~\eqref{e14} are equal to each other up to a coefficient which can be absorbed in the four-dimensional fields ${\overline {\bf 5}^I}$. 
The harmonic representatives of~\eqref{e14} are given by 
\be
\hat{\o}_{1, 2}^{({\overline {\bf 5}^1})} =\hat{\o}_{1, 5}^{({\overline {\bf 5}^2})} =\frac{1}{(1+ |w_3|^2)^2 (1+ |w_4|^2)^2 } d {\overline w}_3 \wedge d {\overline w}_4\;. 
\label{e15}
\ee
The solution for $\hat{\nu}^{({\overline {\bf 5}_I})}$ can be found from~\eqref{e13} using~\eqref{e15} and the explicit formulas for the polynomials $p_1, \dots, p_5$ 
but fortunately we will not need it. 

Finally, we apply the same procedure to $\hat{\nu}^{({\bf 10}^I)}$. From Eq.~\eqref{HL4L5} we see that it is also related to a closed $(0,2)$-form and the system 
of equations describing it is 
\bea
&&
{\overline \pt} \hat{\nu}^{({\bf 10}^I)}= 
p  \hat{\omega}_1^{({\bf 10}^I)} \,,  
\nonumber \\
&&
\overline \pt \hat{\o}_1^{({\bf 10}^I)} = 0\,.    
\label{e16}
\eea
The (column) vectors $\hat{\o}_{1, a}^{({\bf 10}^I)}$ have the following non-vanishing components
\be
\hat{\o}_{1, 1}^{({\bf 10}^1)} \in H^2 ({\cal A}, {\cal N}_1^* \otimes   {\cal L}_5)\;, \quad 
\hat{\o}_{1, 3}^{({\bf 10}^2)} \in H^2 ({\cal A}, {\cal N}_3^* \otimes {\cal L}_4 )\;.
\label{e17}
\ee
Since 
\bea
&&
H^{2} ({\cal A}, {\cal N}^*_1   \otimes  {\cal L}_5) =  H^2 ({\cal A}, {\cal N}_3^* \otimes {\cal L}_4 )
=H^{2} ({\cal A}, {\cal O}_{{\cal A}} (-2, -2, 0, 0, 0, 0))\;, \nonumber \\
&& 
h^{2} ({\cal A}, {\cal O}_{{\cal A}} (-2, -2, 0, 0, 0, 0))=1
\label{e17.0}
\eea
it follows that 
\be
\hat{\o}_{1, 1}^{({\bf 10}^1)} =\hat{\o}_{1, 3}^{({\bf 10}^2)} =\frac{1}{(1+ |w_1|^2)^2 (1+ |w_2|^2)^2 } d {\overline w}_1 \wedge d {\overline w}_2\;. 
\label{e18}
\ee

In the upstairs theory we have two down Yukawa couplings
\be 
\hat{W}=  \hat{\lambda}_{1, I} H^I_{1, 2} {\overline {\bf 5}}_{3,4} {\bf 10}_5 +  \hat{\lambda}_{2, I} H^I_{1, 2} {\overline {\bf 5}}_{3,5} {\bf 10}_4\;, 
\label{e19}
\ee
where $ \hat{\lambda}_{1, I}$ and $ \hat{\lambda}_{2, I} $ are given by 
\bea
&& 
\hat{\lambda}_{1, I} = \frac{1}{ (2 \pi i)^5} \int_{{\mathbb C}^8}\mu \wedge \hat{\nu}^{(H^I)}  \wedge  \hat{\nu}^{(\overline{\bf 5}^1)} \wedge \hat{\nu}^{({\bf 10}^1)}
\wedge {\overline \pt} \Big(\frac{1}{p_1} \Big)\wedge \dots \wedge {\overline \pt} \Big(\frac{1}{p_5} \Big)\;, \nonumber \\
&& 
\hat{\lambda}_{2, I} = \frac{1}{ (2 \pi i)^5} \int_{{\mathbb C}^8}\mu \wedge \hat{\nu}^{(H^I)}  \wedge  \hat{\nu}^{(\overline{\bf 5}^2)} \wedge \hat{\nu}^{({\bf 10}^2)}
\wedge {\overline \pt} \Big(\frac{1}{p_1} \Big)\wedge \dots \wedge {\overline \pt} \Big(\frac{1}{p_5} \Big)\;.
\label{e20}
\eea
To compute $\hat{\lambda}_{1, I}$ and   $\hat{\lambda}_{2 , I}$ we integrate by parts using Eqs.~\eqref{e9}, \eqref{e13}, \eqref{e16}
and the maps $p, q_a$. Fortunately, our analysis simplifies because the total degree of the closed forms $\o_{3, abc}^{(H^I)}$, 
$\o_{1, a}^{({\overline {\bf 5}}^I)}$, $\o_{1, a}^{( {\bf 10}^I)}$ is $4+2+2=8$ which is the dimension of ${\cal A}$. 
On general grounds, after integration by parts we have to obtain the following result
\be 
 \int_{{\mathbb C}^8}\mu \wedge \beta_8
 \label{e21}
 \ee
for some  $(0, 8)$-form $\beta_8$. There is only one possibility to create $\beta_8$ out of $\o_{3, abc}^{(H^I)}$,  $\o_{1, a}^{({\overline {\bf 5}}^I)}$, $\o_{1, a}^{( {\bf 10}^I)}$
and the lower-degree forms arising as partial solutions of Eqs.~\eqref{e9}, \eqref{e13} and~\eqref{e16} which is 
$\beta_8 \sim \o_{3}^{(H^I)} \wedge  \o_{1}^{({\overline {\bf 5}}^I)} \wedge  \o_{1}^{( {\bf 10}^I)}$.
However, one has to be more specific because all these forms carry indices which must be appropriately contracted. 
For the integral~\eqref{e21} to make sense the form $\beta_8$ must take values in  the canonical bundle of ${\cal A}$ 
\be 
{\rm K}_{{\cal A}} \simeq \Lambda^5 {\cal N}^* = {\cal N}^*_1 \otimes  {\cal N}^*_2 \otimes  {\cal N}^*_3 \otimes  {\cal N}^*_4 \otimes  {\cal N}^*_5 \;. 
\label{e22}
\ee
This means that only such combinations of components can appear in $\o_{3}^{(H^I)} \wedge  \o_{1}^{({\overline {\bf 5}}^I)} \wedge  \o_{1}^{( {\bf 10}^I)}$
in which each ${\cal N}_a^*$ appears exactly once. Looking at eqs.~\eqref{e10}, \eqref{e14}, \eqref{e17} we then conclude that 
$\hat{\lambda}_{1, 2}= \hat{\lambda}_{2, 1}=0$. That, the first Higgs particle couples only to ${\overline {\bf 5}}_{3,4} {\bf 10}_5$ and the second 
Higgs particle couples only to ${\overline {\bf 5}}_{3,5} {\bf 10}_4$.\footnote{The vanishing of the couplings $ H^2_{1,2} {\overline {\bf 5}}_{3,4} {\bf 10}_5$
and $ H^1_{1,2} {\overline {\bf 5}}_{3,5} {\bf 10}_4$ is pure geometric and cannot be explained by symmetries of the theory.}
Up to an overall coefficient there is a unique way to build the general expression for $\beta_8$ 
satisfying the above properties. It is given by 
\be 
\beta_8 = \epsilon_{abcde} \o_{3, abc}^{(H)} \wedge \o_{1, d}^{({\overline {\bf 5}})} \wedge \o_{1,e}^{( {\bf 10})}\;, 
\label{e23}
\ee
where it is assumed that $a<b<c$ and $\epsilon_{abcde}$ is totally antisymmetric with $\epsilon_{12345}=-1$. 
The overall coefficient can be fixed by performing a  sample calculation when  $\o_{3}^{(H)}$,  $ \o_{1}^{({\overline {\bf 5}})}$, $  \o_{1}^{( {\bf 10})}$
each has only one component and these components can combine according to Eq.~\eqref{e23}. 
This fixes 
$\beta_8$ in the form~\eqref{e23}. We will not present this calculation in the paper 
because it is rather lengthy and the precise value of the coefficient is not important for our discussion  (as long as it is non-zero). 
Then using Eqs.~\eqref{e10}, \eqref{e14}, \eqref{e17} we find that 
\be 
\hat{\lambda}_{1, 1} = \frac{1}{ (2 \pi i)^5} \int_{{\mathbb C}^8}\mu \wedge \epsilon_{abcde} \o_{3, abc}^{(H^1)} \wedge \o_{1, d}^{({\overline {\bf 5}}^1)} \wedge \o_{1, e}^{( {\bf 10}^1)}
\label{e24}
\ee
for $(a, b, c)= (3, 4, 5)$, $d=2$, $e=1$ and 
\be 
\hat{\lambda}_{2, 2} = \frac{1}{ (2 \pi i)^5} \int_{{\mathbb C}^8}\mu \wedge \epsilon_{abcde} \o_{3, abc}^{(H^2)} \wedge \o_{1, d}^{({\overline {\bf 5}}^2)} \wedge \o_{1,e}^{( {\bf 10}^2)}
\label{e25}
\ee
for $(a, b, c)= (1, 2, 4)$, $d=5$, $e=3$. Substituting now the forms using Eqs.~\eqref{e12}, \eqref{e15}, \eqref{e18} we obtain 
\be 
\hat{\lambda}_{1, 1}  = a_1 y \;, \quad \hat{\lambda}_{2, 2}  = -a_2 y\;, 
\label{e26}
\ee
where 
\bea
&& 
y= \frac{1}{(2 \pi i)^5} {\cal I}_1^4  {\cal I}_2^2\;, \nonumber \\
&& 
{\cal I}_1= \int_{{\mathbb C}^8} \frac{dw \wedge d {\overline w}}{ (1+ |w|^2)^2}\;, \quad 
{\cal I}_2= \int_{{\mathbb C}^8} \frac{dw_1 \wedge d {\overline w}_1 \wedge dw_2 \wedge d {\overline w}_2}{ (1+ |w_1|^2 + |w_2|^2 )^3}\;. 
\label{e27}
\eea
Evaluating the integrals gives\footnote{The integral  $(2 \pi i)^{-1}{\cal I}_1$ is just the integral of the K\"ahler form $J$ over ${\mathbb P}^1$ 
which is normalized to $1$. The integral  $(2 \pi i)^{-2}{\cal I}_2$ is the integral of $J \wedge J$ over ${\mathbb P}^2$ which is also normalized to $1$.}
\be 
{\cal I}_1 = 2 \pi i\;, \quad {\cal I}_2 = (2 \pi i)^2\;, \quad y = (2 \pi i )^3\;. 
\label{e28}
\ee

The down Yukawa coupling in the upstairs theory is then given by 
\be 
\hat{W}=  y (a_1 H^1_{1, 2} {\overline {\bf 5}}_{3,4} {\bf 10}_5 -  a_2 H^2_{1, 2} {\overline {\bf 5}}_{3,5} {\bf 10}_4)\;.
\label{e29}
\ee
This formula is similar to Eq.~\eqref{Wex} except we have two Higgs fields. 
However, when we mod out by the action of ${\mathbb Z}_2$ to go to the 
Standard Model only one Higgs field will survive the projection. To bring~\eqref{e29} to the form~\eqref{Wex} lets us eliminate the Higgs field 
which will not descent to the Standard Model. 
Using the properties of the ${\mathbb Z}_2$  action in~\eqref{ctrans}, \eqref{Ntrans} and 
Eqs.~\eqref{e10}, \eqref{e12} it follows that the ${\mathbb Z}_2$ action interchanges the forms $\o_{3, 345}^{(H^1)}$ and $\o_{3, 124}^{(H^2)}$ or, equivalently, 
it interchanges $a_1$ and $a_2$. Let us recall from Eq.~\eqref{chiSM}  that the Higgs field has the charge $\chi_2$ under the discrete symmetry $\Gamma$. 
For $\Gamma = {\mathbb Z}_2$ it is easy to realize that $\chi_2$ must be non-trivial, that is the Higgs field is odd under  ${\mathbb Z}_2$. 
This means that the appropriate linear combination of the forms in~\eqref{e12} which  will descend to the downstairs Calabi-Yau threefold $X$ is their  difference $\sim a_1 -a_2$. 
Similarly, the appropriate linear combination of the Higgs fields which will descent to the Standard Model is $H_{1,2}= H^1_{1, 2}- H^2_{1,2}$. 
Ignoring the other linear combination $H^1_{1, 2}+ H^2_{1,2}$ and absorbing $a_1-a_2$ into $H_{1,2}$ we obtain 
\be 
\hat{W}=  y H_{1,2} {\overline {\bf 5}}_{3,4} {\bf 10}_5 - y H_{1,2} {\overline {\bf 5}}_{3,5} {\bf 10}_4\;. 
\label{e30}
\ee
Thus, we obtain precisely Eq.~\eqref{Wex} where $y'= -y$ and $y$ is given by Eq.~\eqref{e28}. As was discussed before this leads to Yukawa unification for one family in the downstairs theory.


\section{Conclusion}\label{sec:concl}
In this paper, we have discussed Yukawa unification in the context of heterotic Calabi-Yau models based on the standard, two-step construction. This involves a non-flat gauge bundle, which breaks $E_8$ to a more standard GUT group, in the first step. The second step is to introduce a Wilson line on a quotient of the original manifold, breaking the gauge group to the standard model. As reviewed in the introduction, models of this kind are the only ones in the context of smooth, K\"ahler, heterotic compactifications that are capable of producing a realistic low-energy spectrum. Our main question has been whether such models can ever lead to Yukawa unification similar to that seen in traditional field theory GUTs.

We have provided a detailed analysis of the fact \cite{Green:1987mn} that such unification is never enforced by the underlying GUT symmetry, at least for the two main GUT groups $SU(5)$ and $SO(10)$ on which we have focused. The reason for this can be easily understood qualitatively. The standard model and the underlying GUT theory are related by a quotient with a discrete symmetry $\Gamma$. In order to obtain three standard model families the GUT theory requires $3|\Gamma|$ families and it has, hence, larger Yukawa matrices of size $(3|\Gamma|)\times (3|\Gamma|)$. The standard model Yukawa matrices always originate from different parts of the larger upstairs Yukawa matrix. Hence, the GUT group never enforces Yukawa unification for such models.

Additional symmetries in the GUT theory can, however, lead to relations between the upstairs Yukawa couplings which, in turn, may translate into Yukawa unification in the downstairs model. We have studied the possibility that the discrete symmetry $\Gamma$, together with possible additional $U(1)$ gauge factors, can play this role. It turns out that these symmetries do not lead to unification if they commute. In contrast, we have presented two scenarios in the context of heterotic line bundle models where the discrete groups $\Gamma=\mathbb{Z}_2, \mathbb{Z}_3$ do not commute with some of the high energy $U(1)$ symmetries, and where (full or partial) Yukawa unification does occur. In particular, it is possible to unify Yukawa couplings for one family but not the others.

Finally, as a proof of existence, we have presented an explicit heterotic line bundle model based on $SU(5)$, where this scenario is realized for $\Gamma=\mathbb{Z}_2$. It is clear that such models are quite rare and difficult to find. 

In this paper, we have focused on obvious sources of additional symmetries, namely the discrete symmetry $\Gamma$ and additional $U(1)$ factors which can originate from split bundles. Further discrete symmetries might be available in specific models and might also result in complete or partial Yukawa unification. 


\section*{Acknowledgements}
The work of E.I.B. was supported by the ARC Future Fellowship FT120100466 and in part by the ARC Discovery project DP140103925.
A.L. is partially supported by the EPSRC network grant EP/N007158/1 and by the STFC grant~ST/L000474/1. The work of J.G. is supported by NSF grant PHY-1417316.
E.I.B. and A.C.~would like to thank the Physics department at the University of Oxford for hospitality, where part of this work has been carried out.
A.L. would like to thank Stuart Raby and Fabian Ruehle  for helpful discussions of Yukawa unification in orbifold models which have partially inspired this work.


\bibliography{bibfile}{}
\bibliographystyle{utcaps}

\end{document}